\documentclass[aps,pre,twocolumn,superscriptaddress]{revtex4-1}

\usepackage[english]{babel}
\usepackage[utf8x]{inputenc}
\usepackage{amsmath,amsfonts,amssymb,amsbsy,eucal,dsfont}
\usepackage{natbib}
\usepackage{color}
\usepackage{graphicx}
\usepackage{rotating}
\usepackage{longtable}
\begin{document}
\title{Electrokinetic and hydrodynamic properties of charged-particles systems:\\
From small electrolyte ions to large colloids}

\author{G. N\"agele}
\email[]{g.naegele@fz-juelich.de}
\affiliation{Institute of Complex Systems (ICS-3),
Research Centre J\"ulich,
52425 J\"ulich,
Germany,
}

\author{M. Heinen}
\affiliation{Institut f\"{u}r Theoretische Physik II,
Weiche Materie,
Heinrich-Heine-Universit\"{a}t D\"{u}sseldorf,
40225 D\"{u}sseldorf,
Germany}

\author{A.J. Banchio}
\affiliation{FaMAF, Universidad Nacional de C\'{o}rdoba, and
IFEG-CONICET, Ciudad Universitaria, X5000HUA C\'{o}rdoba,
Argentina}

\author{C. Contreras - Aburto}
\affiliation{Divisi\'{o}n de Ciencias e Ingenier\'{\i}as,
Universidad de Guanajuato Campus Le\'{o}n,
37150 Le\'{o}n,
M\'{e}xico}

\begin{abstract}
Dynamic processes in dispersions of charged spherical particles are
of importance both in fundamental science, and in technical and bio-medical
applications. There exists a large variety of charged-particles systems,
ranging from nanometer-sized electrolyte ions
to micron-sized charge-stabilized colloids. We review recent advances
in theoretical methods for the calculation of linear transport coefficients
in concentrated particulate systems, with the focus on hydrodynamic
interactions and electrokinetic effects. Considered transport properties are the dispersion
viscosity, self- and collective diffusion coefficients, sedimentation
coefficients, and electrophoretic mobilities and conductivities of
ionic particle species in an external electric field. Advances by
our group are also discussed, including a novel mode-coupling-theory
method for conduction-diffusion and viscoelastic properties of strong
electrolyte solutions. Furthermore, results are presented for dispersions
of solvent-permeable particles, and particles with non-zero hydrodynamic
surface slip. The concentration-dependent swelling of ionic microgels
is discussed, as well as a far-reaching dynamic scaling behavior relating
colloidal long- to short-time dynamics.
\end{abstract}
\maketitle

\section{Introduction\label{sec:Introduction}}

Dispersions of charged globular particles undergoing correlated Brownian
motions are ubiquitously found with a broad range of particle sizes,
ranging from electrolyte solutions through solutions of nanometer-sized
globular proteins to dispersions of micron-sized charge-stabilized
colloidal spheres.

Charge-stabilized colloid particles are encountered in a rich variety
in chemical industry, biology and food science. Composite and aspherical
colloidal particles have attracted increasing attention in soft matter
science. These classes of colloids include suspensions of rods, discs,
core-shell particles with a solid core and surrounding polyelectrolyte
brush layer, star polymers and ionic microgels, to name only a few
examples. The stimuli-dependent size and biocompatibility of microgels
such as poly (N-isopropylacrylamide) (PNiPAm) allows for their use in
biomedical applications including drug delivery. Experimentally well-studied
examples of globular proteins are, e.g., bovine serum albumin (BSA),
lyzoyzme, and apoferritin \cite{2005JChPh.123e4708G,2012SMat....8.1404H,RoosenRunge:2011,Goegelein:2008,Egelhaaf2004,Goegelein:2012}.
If dispersed in water, the proteins are moderately charged, by an
amount depending on temperature, pH value, salinity, and protein concentration.
A quantitative understanding of the dynamics in concentrated solutions
of interacting proteins is of importance, e.g., to the evaluation
of cellular functions. Conduction-diffusion and viscoelastic transport
coefficients of electrolyte solutions are of relevance in electrochemistry,
geology, energy research, and biology \cite{ISI:A1991GV01300022,Nonner:1999hx}.
They play an important role in many industrial processes including
waste water treatment and ion exchange applications. A thorough theoretical
understanding of electrolyte transport properties of non-dilute solutions
has still not been achieved, in spite of the large body of accumulated
empirical data \cite{Jenkins:1995,WOLYNES:1980}.

While on first sight electrolyte ions, globular proteins and spherical
charged colloids appear to be very different, from a simplifying theoretical
viewpoint they can be all treated as (uniformly) charged Brownian
spheres, interacting by Coulomb plus excluded volume forces, and immersed
in a structureless Newtonian solvent characterized by the dielectric
constant $\epsilon$ and the shear viscosity $\eta_{0}$. In this
primitive model (PM) type picture, the dynamics of all ionic species,
i.e. colloid and protein macroions as well as electrolyte microions,
is taken to be overdamped, with the configurational evolution of all
ions described by the many-particle generalized Smoluchowski equation
(GSmE) in combination with the low-Reynolds number creeping flow equation
for the solvent. Such a simplifying description makes good sense also
for electrolyte solutions \cite{Barthel1998}, even though ion-solvent
specific effects on the molecular level \cite{Falkenhagen:1971vq},
commonly addressed using concepts such as hydration shells, local
solvent polarization by ion electric fields, and structure-breaking
and structure-making ion properties, are \emph{per se} not accounted
for in the PM-GSmE model. However, ion hydration shells can be approximately
accounted for in a structureless solvent model by using mixed slip-stick
hydrodynamic boundary conditions (BCs) on the PM microion surfaces,
and by allowing for a certain solvent permeability.

The dynamics of macroions and microions is determined by the interplay
of direct electrosteric forces and indirect, solvent-mediated hydrodynamic
interactions (HIs). The latter are long-ranged and in general of many-body
nature \cite{NaegeleVarena2013}. This causes challenging problems
in theoretical and computer simulation studies of Brownian ion systems.
The dynamics of colloidal macroions much larger than electrolyte and
surface-released counterions is therefore frequently described using
the one-component macroion fluid model (OMF). In this simplifying
model, only the microion-dressed colloidal particles are considered
which interact for non-overlap distances by an effective screened
Coulomb pair potential.  The averaged microion degrees of freedom
enter into the OMF description only through effective colloidal charge
number and electrostatic screening parameters, which in general are
dependent on the thermodynamic state. The simplicity of the OMF allows
for a quantitative consideration of the strong colloid-colloid HIs
in dense charge stabilized suspensions. However, it disregards electrokinetic
effects on the colloid dynamics, arising in particular from the non-instantaneous
dynamic response, to internal or external perturbations, of the microion
cloud surrounding each colloidal macroion. This so-called microion
relaxation mechanism is most influential for low colloid concentrations
and smaller macroion sizes, and when the Debye effective screening
length is comparable to the macroion radius. An additional electrokinetic
mechanism of purely hydrodynamic origin becomes operative when an
electric field is applied (electrophoresis). This so-called electrophoretic
mechanism describes the slowing influence on the field-induced macroion
migration owing to the hydrodynamic coupling with surrounding counter-
and coions.

Colloidal electrokinetic effects have been studied in the past mainly
using the standard electrokinetic theory (SET) approach where the
microions are described as continuous local charge densities coupled
to the solvent creeping flow, with advection-diffusion type Nernst-Planck
constitutive equations for the microion fluxes \cite{Ohshima:2006tk,BookMaslijay:2006}.
The SET equations are commonly linearized with respect to the driving
field. In this mean-field type approach, microion correlations are
disregarded which are important when the macroions are small such
as in the case of proteins, or when the macroions are strongly charged
and non-monovalent counterions are present. Microion correlation
effects have been included approximately in recent extensions of the
SET equations \cite{Bazant200948,FLM:6774480,C1CP22359C,LopezGarcia:2011}.
In the extension by Lozada-Cassou and collaborators \cite{LozadaCassou:1999,Lozada2001,Lozada2001Erratum},
microion finite sizes are incorporated through a non-ideal contribution
to the microionic electrochemical potential, evaluated to first order
in the external electric field using the hybrid HNC-MSA equilibrium
ionic pair distribution functions for a restricted PM electrolyte
surrounding a spherical macroion. The macroion zeta potential at the
slipping surface is identified with the equilibrium mean electrostatic
potential at the closest approach distance to the microion (see also
\cite{ManzanillaGranados:2011}). A possible sign reversal of the
electrophoretic mobility is hereby directly linked to a corresponding
colloidal charge reversal. In the approach by Lozada-Cassou \emph{et
al.}, the microions are dynamically still treated as continuous charge
density fields. The reduced screening ability of non-zero-sized microions
reduces the relaxation effect, giving rise to an enlarged macroion
electrophoretic mobility at large zeta potential values.

While developed originally for a single colloidal particle in an electrolyte
solution, the SET approach has been extended over the years to concentrated
suspensions, mostly on basis of simplifying cell models \cite{Carrique2001157,Carrique200543,doi:10.1021/jp054969r,Ding2001180,Kozak1989497,Kozak1989166,Levine1974520,Ohshima1997481,Ohshima2000140,Ohshima:2006tk,Shilov1981,Zholkovskij2007279}.
These cell model SET extensions, although being quite successful in
predicting general trends in electrophoresis and sedimentation, are
of an \emph{ad hoc} nature and do not properly account for macroion-macroion
correlations arising from overlapping colloidal electric double layers
(EDLs) and colloid-colloid HIs. This is also reflected in the still
ongoing discussion about the appropriate electric and hydrodynamic
BCs at the outer cell boundary (see, e.g. \cite{CJCE:CJCE5450850517}).
The SET cell models predict the electrophoretic mobility to decrease
in general with increasing volume fraction, also for low-polar solvents
\cite{Vissers:2011}, and with increasing overlap of the colloidal
EDLs. In most calculations, either a constant surface potential or
surface charge density have been assumed, but chemically charge-regulated
colloidal surfaces are also considered \cite{Hsu:2000}. Moreover,
cell model dc and ac electrophoretic calculations for core-shell spheres
with microion-penetrable shells have been made \cite{LopezGarcia:2006,Ahuali:2009}.

There exist also two-colloid extensions of the SET approach which
have been used, e.g., for determining the concentration dependence
of the electrophoretic mobility in semi-dilute suspensions \cite{EnnisWhite:1997,EnnisWhite_Corrigendum:1997,ShugaiCarnie:1997}.
This two-colloid approach gives in particular the correct solvent
backflow factor $\left[1-1.5\;\!\phi+{\cal O}\left(\phi^{2}\right)\right]$
with the colloid volume fraction $\phi$, multiplying the single-colloid
Smoluchowski electrophoretic mobility of monodisperse and non-conducting
colloidal spheres with ultrathin EDLs. The electrophoretic mobility
and the flow behavior of suspensions of interacting colloidal charged
spheres can be determined from the measured power spectrum using low-angle
super-heterodyne Doppler velocimetry. Experimental results by this
method, and an outline of the underlying light scattering theory,
are given in \cite{Palberg:2012}.

We point out that different from colloidal systems, all ions in electrolyte
solutions are of comparable size, charge, and mobility. Therefore,
a full PM-GSmE description of the microion electrokinetics is required
for non-dilute systems with all ions treated individually as dynamic
entities.

In this work, we review recent advances in the theoretical understanding
of linear diffusion-convection and rheological transport properties
characterizing dispersions of globular charged Brownian particles.
The review encompasses a broad range of properties, including microionic
conductivities and electrophoretic mobilities, high-frequency and
steady-state viscosities, generalized sedimentation coefficients,
and self- and collective diffusion coefficients.

Theoretical methods for the calculation of conduction-diffusion coefficients
and the viscosity of strong electrolyte solutions are reviewed in
Sec. \ref{sec:Electrolytes}. The section includes new results which
we have obtained using a simplified mode-coupling theory (MCT) method
where ion-ion HIs are properly accounted for not only in the short-time
response, but also in the microion clouds relaxation contribution.
In Sec. \ref{sec:Crowded-globular-protein}, the dynamics in solutions
of globular proteins is addressed. We demonstrate that theoretical
methods developed originally for colloids can be successfully applied
to crowded protein solutions such as BSA. Sec. \ref{sec:Charge-stabilized-colloidal-suspensions}
reports on recent progress in the understanding of transport properties
of concentrated suspensions of charge-stabilized colloidal particles.
In particular, the dynamic behavior of solvent-permeable particles,
and of ionic microgels penetrable by surrounding counterions is discussed.
Our conclusions are contained in Sec. \ref{sec:Conclusions}.
A list of abbreviations is included following the acknowledgements.

\section{Electrolyte solutions \label{sec:Electrolytes}}

We consider here strong electrolyte solutions where the salt solute
is fully dissociated. At total electrolyte concentrations $n_{T}$
lower than about $0.01$ M, electrolyte ions can be treated as pointlike,
and their Coulomb interactions give rise to the peculiar square-root
in concentration dependence of electrolyte transport properties. This
concentration dependence is the hallmark of the Debye-Falkenhagen-Onsager-Fuoss
(DFOF) limiting law expressions for the transport coefficients characterizing
electrolyte conductivity and electrophoresis \cite{Onsager:1932ux,Onsager:1957vv},
self-diffusion \cite{Onsager:1945we}, and viscosity \cite{Onsager:1932ux,Falkenhagen:1932us,Onsager:1957vv}.
The limiting law expressions have been derived using a continuum model
for the solvent, with the ions treated as pointlike Brownian particles
described by equilibrium pair distribution functions on the linear
Debye-H\"uckel (DH) theory level.

Various routes have been followed in the past for calculating conduction-diffusion
and rheological properties of non-dilute electrolytes where the excluded
volume of the ions needs to be considered. Falkenhagen \cite{Falkenhagen:1971vq} and Ebeling
\emph{et al}. \cite{EBELING:1978uo,KREMP:1983vu} have extended
the DFOF continuity equations approach to finite ion sizes. The relaxation
mechanism contribution to the conductivity is deduced in their approach
from averaging the electrostatic force experienced by a central ion
using the perturbed ionic pair distribution functions.

A considerable improvement over the DFOF theory was obtained by Bernard,
Turq, Blum, Dufreche and collaborators in a series of publications
\cite{Bernard:1992tj,DurandVidal:1996tg,DurandVidal:1996vn,Dufreche:2005fj,Roger:2009gv}
where the DFOF approach has been combined (mostly) with the analytic
mean-spherical approximation (MSA) solution \cite{Blum:1977ep,HIROIKE:1977wm}
for the ionic pair distribution functions. They obtained results for
the steady-state ion conductivity \cite{Bernard:1992tj,DurandVidal:1996vn,Roger:2009gv},
the ion self-diffusion coefficients \cite{Bernard:1992we}, and the
mutual (chemical) diffusion coefficient \cite{Dufreche:2002wk,DufrecheBrownian:2005}
(see also \cite{Felderhof:2003ck,Dufreche:2003ku}). The in general
non-negligible influence of the ion-ion HIs on the ion cloud relaxation
mechanism is disregarded in their treatment, except for the special
case of ionic self-diffusion \cite{Dufreche:2008fla}. Moreover, the
electrolyte viscosity has not been considered in their works

In recent works, Chandra, Bagchi and collaborators have combined MCT
and dynamic density functional theory (DDFT) arguments to derive expressions
for the molar conductivity \cite{Chandra:1999um,Chandra:2000kj,Dufreche:2005fj}
and viscosity \cite{Chandra:2000um} of electrolyte solutions. The
ion excluded volumes are incorporated in their hybrid method using
Attard's generalization \cite{ATTARD:1993vr} of the DH pair distribution
functions. However, the influence of the ion-ion HIs on the viscosity,
and on the relaxation mechanism part of the conductivity, are disregarded
in their MCT-DDFT treatment. In related work, Dufreche \emph{et al}.
\cite{Dufreche:2008fla} have combined MCT and DDFT arguments with
Kirkwood's friction formula for electrolyte friction to calculate
the ion self-diffusion coefficients and velocity autocorrelation functions
in a binary electrolyte solution. The finite ion sizes in this approach
to self-diffusion are accounted for in MSA, and the inter-ion HIs
are treated on the point-particle (Oseen) level of description.
In \cite{Dufreche:2008fla}, however, the effect of dynamic cross
correlations in the intermediate scattering functions input is disregarded.

In a series of papers, using linear response theory we have developed
a unifying MCT method for calculating linear
conduction-diffusion \cite{Contreras_I_Paper:2013,Contreras_II_Paper:2013}
and viscoelastic \cite{Contreras2012visc,Contreras_II_Paper:2013}
properties of non-dilute strong electrolyte solutions. This method
builds on earlier work where a general MCT for the dynamic structure
factor of Brownian particle mixtures with HIs has been developed \cite{Nagele:1998tp,Nagele:1998tt,Nagele:1999vh}.
Our statistical mechanical theory of electrolyte transport on the
PM-GSmE level includes the influence of the solvent-mediated ion-ion
HIs not only in the short-time response, but also in the relaxation
mechanism contribution. It provides hereby a complete description
of steady-state transport coefficients. This differentiates our method
from earlier theoretical work where HIs were only incompletely considered,
or where severe approximations such as the Nernst-Einstein relation
between ion self-diffusion and conductivity have been invoked. Using
a simplifying solution scheme, referred to as the MCT-HIs approach,
easy-to-apply semi-analytic transport coefficient expressions have
been derived and evaluated for an aqueous $1:1$ strong electrolyte solution which is an overall electroneutral binary
solution of monovalent cations and anions in water.
The predicted coefficients for stick hydrodynamic BC on the PM ion surfaces
agree well with experimental data for the electrolyte conductivity
and viscosity, for ion concentrations extending even up to two molar.
To analyze the dynamic influence of ion hydration shells, \cite{Contreras_II_Paper:2013}
includes a discussion on the significance of mixed stick-slip hydrodynamic
surface boundary conditions, and on the effect of solvent permeability.

\begin{figure}
\centering
\includegraphics[width=1\columnwidth]{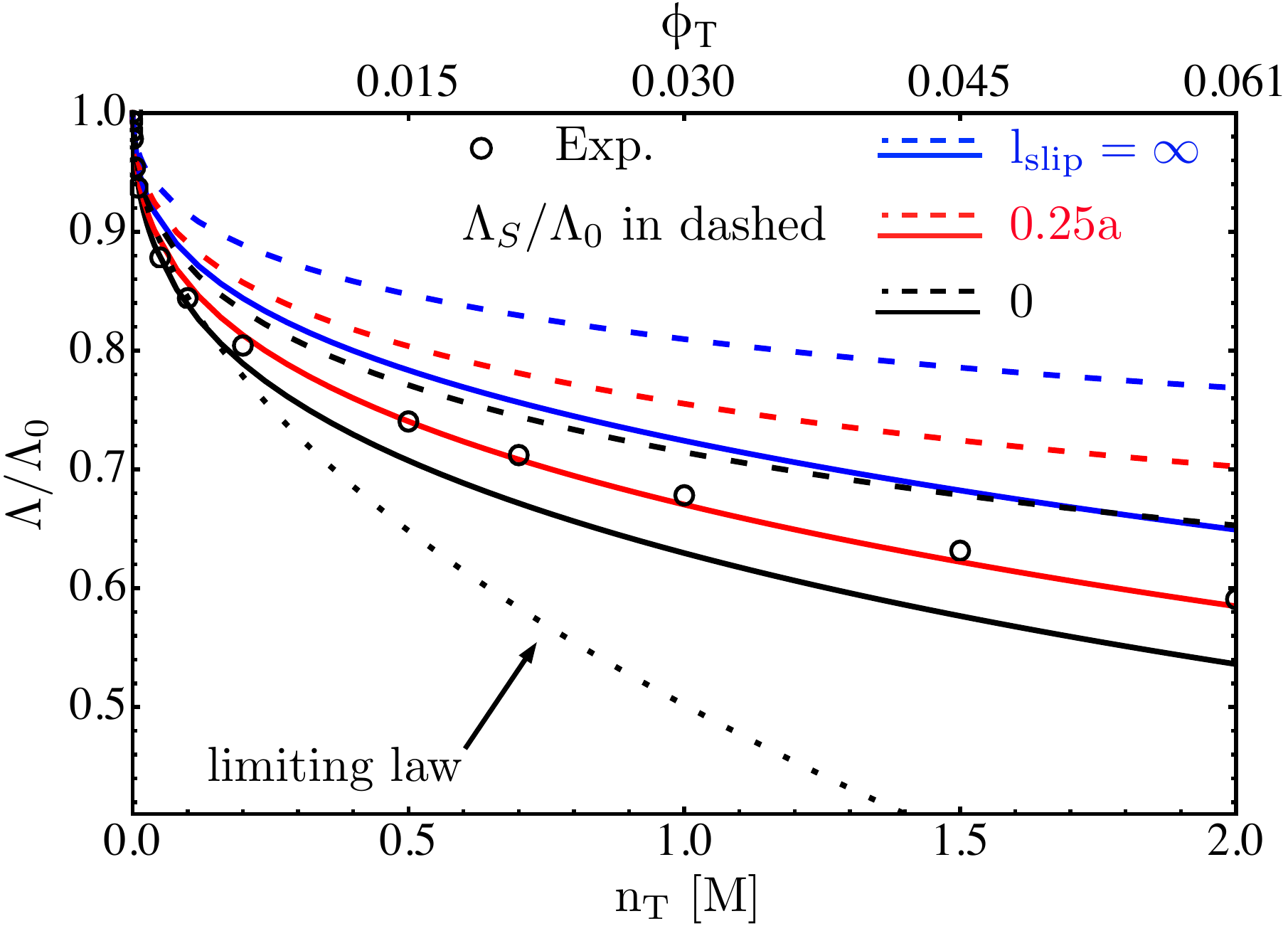}
\caption{MCT-HIs based theoretical predictions
for the normalized molar conductivity, $\Lambda$, of an aqueous $1:1$
electrolyte at $T=25^{\circ}\mbox{C}$ (solid lines), for three hydrodynamic
slip lengths as indicated. The corresponding short-time conductivity
contributions, $\Lambda_{S}$, are shown as dashed lines. The hydrodynamic
mean ion diameter, $\sigma = 2a = 4.58  \text{\AA}$, of hydrated $Na^{+}$
and $Cl^{-}$ ions was used in the calculations. Open circles: experimental
data for the molar conductivity of $NaCl$ dissociated in water, taken
from \cite{Miller1966}. Dotted line: DFOF limiting law result. The
lower (upper) horizontal scale is for the total ion concentration
$n_{T}$ (total ion volume fraction $\phi_{T}$).
}
\label{fig:NaCl-Conductivity}
\end{figure}

\begin{figure}
\begin{center}
\resizebox{1.0\columnwidth}{!}{
\includegraphics{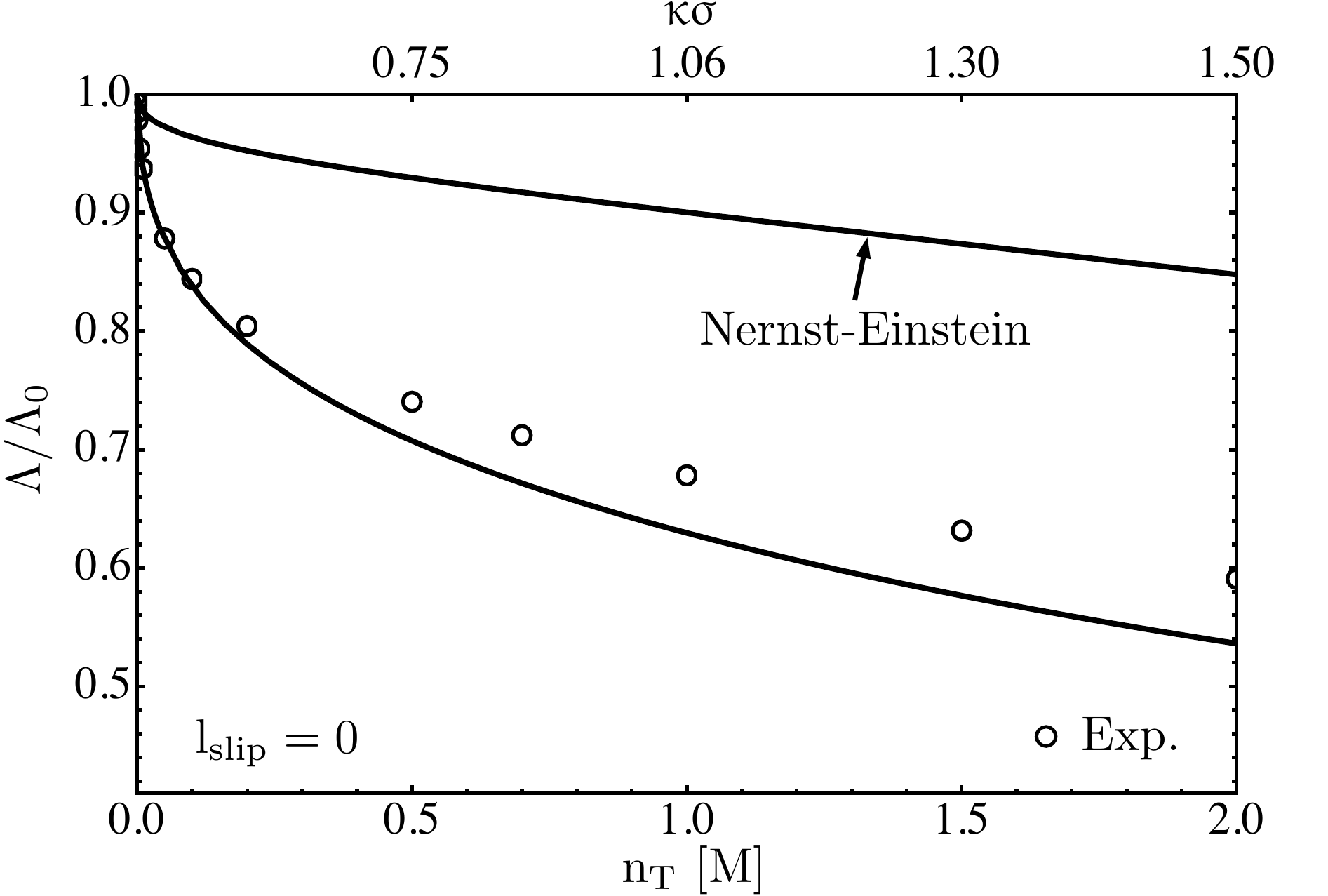}}
\end{center}
\vspace{-1em}
\caption{Accuracy check of the approximate Nernst-Einstein
relation, $\Lambda/\Lambda_{0}\approx d_{L}/d_{0}$, for the reduced
molar conductivity of a symmetric aqueous $1:1$ electrolyte. Solid
lines are simplified MCT-HIs results for $d_{L}/d_{0}$ (upper curve)
and $\Lambda/\Lambda_{0}$ (lower curve), respectively, for zero slip
length. All other system parameters are as in Fig. \ref{fig:NaCl-Conductivity}.
Open circles: experimental conductivity data for $NaCl$ in water
at $T=25^{\circ}\mbox{C}$ \cite{Miller1966}. }
\label{fig:NE-Relation-Test}
\end{figure}

\begin{figure}
\begin{center}
\resizebox{1.0\columnwidth}{!}{
\includegraphics{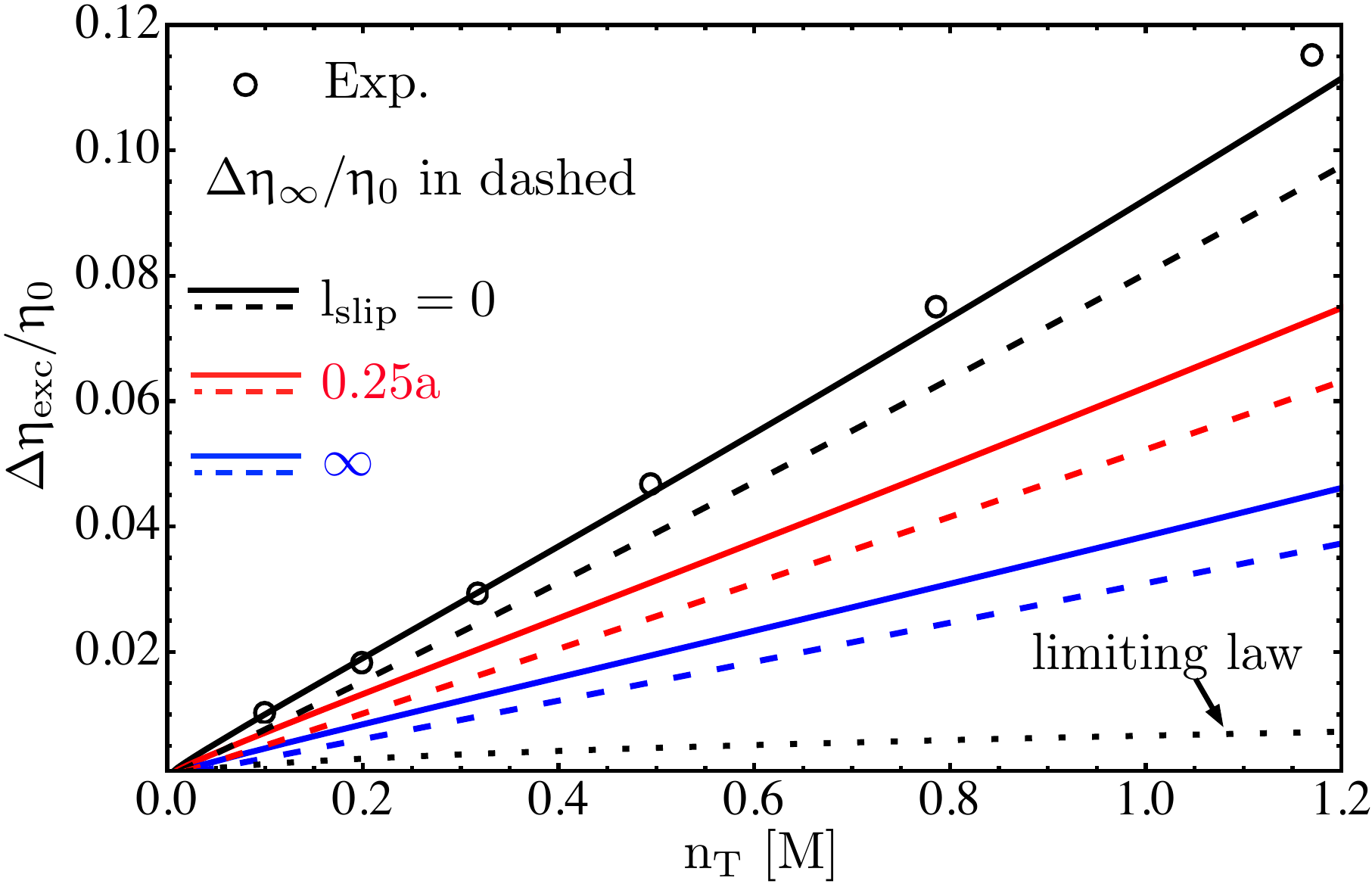}}
\end{center}
\vspace{-1em}
\caption{MCT-HIs based theoretical predictions for
the concentration dependence of the reduced excess electrolyte viscosity,
$\Delta\eta_{exc}/\eta_{0}$, of an aqueous $1:1$ electrolyte at
$T=25^{\circ}\mbox{C}$. Three hydrodynamic slip lengths are considered
as indicated (solid lines), and $\sigma=2a=4.58\text{\AA}$ is used for
the mean ion diameter. The corresponding reduced short-time viscosity
contributions, $\Delta\eta_{\infty}$, are shown as dashed lines.
Open circles: experimental data for the excess viscosity of $NaCl$
in water, taken from \cite{OUT:1980up}. Dotted line: DFOF limiting
law result (see \cite{Contreras2012visc}).}
\label{fig:NaCl-Viscosity}
\end{figure}

A thorough description of the MCT-HIs method with numerous numerical
results is contained in \cite{Contreras2012visc,Contreras_I_Paper:2013,Contreras_II_Paper:2013}.
To illustrate the method, Figs. \ref{fig:NaCl-Conductivity} - \ref{fig:NaCl-Viscosity}
display additional results not included in the aforementioned papers.
In Fig. \ref{fig:NaCl-Conductivity}, the MSA-HIs prediction of the
molar conductivity $\Lambda$ of an aqueous $1:1$ strong electrolyte
solution is compared with experimental data for a $NaCl$ solution.
The quantity $\Lambda$ is the mean electric current density of ions
per unit applied electric field strength, and per mol of salt unit.
Results for mixed slip-stick BCs with three different hydrodynamic slip lengths, $l_{slip}$,
are compared to each other. Here, $l_{slip}$ is the distance into the interior of a PM ion
for which the linear near-surface flow extrapolates to zero.
The experimental conductivity data are
overall well reproduced using the standard stick BC for which $l_{slip}=0$,
even though they are underestimated to some extent at large $n_{T}$.
The agreement between theory and experiment becomes excellent for
$l_{slip}=a/4$, a slip length compatible with the presence of ion
hydration shells which are expected to cause some hydrodynamic slip.
The slip length of an ion depends on the molecular structure and width
of its hydration shell, and for a thicker shell also on the form of
the local flow field which differs in character for electrophoresis
and shear flow without external electric field. Fig. \ref{fig:NaCl-Conductivity}
depicts additionally theoretical results for the short-time (i.e.
electrophoretic mechanism) part, $\Lambda_{S}$, of the conductivity,
normalized by the electrolyte conductivity $\Lambda_{0}$ at infinite
dilution. At larger $n_{T}$, the reduction of the conductivity by
the relaxation mechanism becomes comparatively large to that caused
by the short-time electrophoretic mechanism. As shown in \cite{Contreras_II_Paper:2013},
both the short-time and relaxation parts of $\Lambda$, and of the
viscosity, are considerably affected by the HIs. At very low concentrations,
the DFOF limiting law result for the conductivity is recovered by
the MCT-HIs scheme.


Fig. \ref{fig:NE-Relation-Test} includes a MCT-HIs based analysis
of the approximate Nernst-Einstein (NE) relation which links the conductivity
to the ionic self-diffusion coefficients. For a symmetric $1:1$ electrolyte,
the NE relation simply reads $\Lambda/\Lambda_{0}\approx d_{L}/d_{0}$,
where $d_{L}$ is the long-time self-diffusion coefficient of the
equally sized and equally valent anions and cations with common single-ion
diffusion coefficient $d_{0}$. Owing to the neglected ion velocity
cross correlations in the NE relation, the conductivity is severely
overestimated by this relation.

Similar to the conductivity, the electrolyte viscosity $\eta$ in
excess to the solvent viscosity $\eta_{0}$, is given by the sum \cite{Contreras2012visc},

\[
\Delta\eta_{exc}\;\!=\;\!\eta-\eta_{0}\;\!=\;\!\Delta\eta_{\infty}+\Delta\eta\,\,,
\]
of a short-time (i.e. high-frequency) part $\Delta\eta_{\infty}$
and a shear-stress relaxation part $\Delta\eta$. The short-time part
is of purely hydrodynamic origin and vanishes for point particles
or when HIs are disregarded. Fig. \ref{fig:NaCl-Viscosity} includes
our MCT-HIs predictions for the excess viscosity of an aqueous $1:1$
electrolyte for values of the slip length as in Fig. \ref{fig:NaCl-Conductivity},
in comparison with the measured excess viscosity of $NaCl$ in water.
The experimental data are well described by the MCT-HIs result when the
standard stick BC is used. Usage of the slip length $l_{slip}=a/4$
results however in a significant underestimation of the viscosity.
This can be at least partially attributed to the fact that the mean
local hydrodynamic environment of particles in shear flow is qualitatively
different from the local environment in electrophoresis, with
consequently different hydrodynamic slip lengths \cite{Contreras_I_Paper:2013}.
The strongest contribution to the steady-state viscosity is caused
by the short-time viscosity part $\Delta\eta_{\infty}$. The limiting
law regime, characterized by the $\sqrt{n_{T}}$ dependence of $\Delta\eta_{exc}$,
is reached for very small concentrations only.

\section{Crowded globular protein solutions \label{sec:Crowded-globular-protein} }

\begin{figure}
\begin{center}
\resizebox{1.0\columnwidth}{!}{
\includegraphics{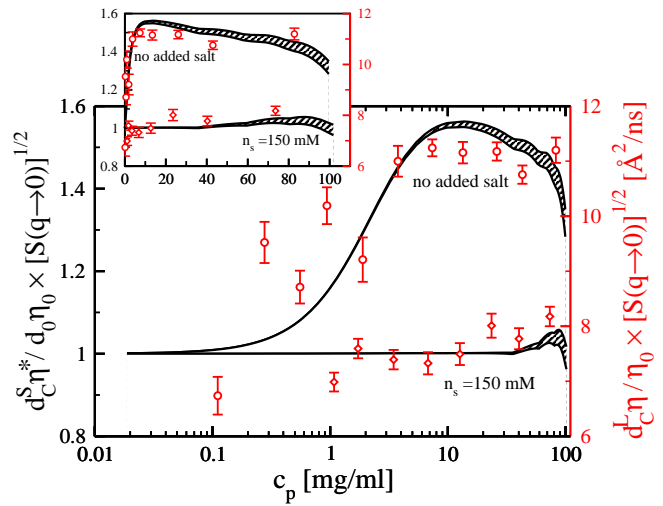}}
\end{center}
\vspace{-1em}
\caption{Experimental-theoretical test of the long-time
and short-time KD-GSE relation with $\eta^{*}=\eta$ and $\eta^{*}=\eta_{\infty}$,
respectively. Results for aqueous BSA solutions without added salt
(upper data sets), and with $150$ mM of added NaCl (lower data sets)
are shown in their dependence on the protein concentration $c_{p}$.
Red symbols: combination of $d_{C}^{L}$ from DLS, $\eta/\eta_{0}$
from suspended couette rheometry, and $S(q\to0)$ from static light
scattering. Black lines: Theoretical results combining $d_{C}^{S}\approx d_{C}^{L}$
and $\eta_{\infty}$, both calculated using the self-part corrected
$\delta\gamma$ scheme \cite{Heinen:2011if}, with $S(q\to0)$ obtained
from the MPB-RMSA scheme. For the long-time KD-GSE relation, $\eta=\eta_{\infty}+\Delta\eta$
has been used, with $\Delta\eta$ calculated using the MCT-HIs scheme.
Lower boundaries of the theoretical curves correspond to the short-time
KD-GSE relation, upper boundaries to the long-time version. The effective
protein diameter is $\sigma=7.40$ nm. Figure taken from \cite{2012SMat....8.1404H}.}
\label{fig:BSA-GSE}
\end{figure}

There have been various attempts in the past to describe the collective
diffusion in globular protein solutions using a simplifying colloid-type
effective sphere model with Derjaguin-Landau-Verwey-Overbeek (DLVO)-like pair interactions. However,
calculations of the collective diffusion coefficient $d_{C}$ based
on this model have been either quite approximate, in particular regarding
the treatment of the protein HIs \cite{BowenMongruel:1998,Yu:2005},
or restricted to the first-order correction in the volume fraction
\cite{PrisenOdijk:2007}. In a recent experimental-theoretical study,
dynamic light scattering measurements of $d_{C}$ and rheometric measurements
of the steady-state viscosity $\eta$ of crowded BSA solutions where
compared with theoretical calculations based on an analytically treatable
spheroid model of BSA with isotropic screened Coulomb interactions.
For calculating the dynamic properties, easy-to-implement theoretical
methods were used which account for the HIs \cite{Heinen:2011if}.
The only input required by these methods is the colloidal static structure
factor $S(q)$, obtained in \cite{2012SMat....8.1404H} using the
newly developed so-called MPB-RMSA integral equation scheme \cite{Heinen_MPBRMSA:2011,Heinen_MPBRMSA_Corrigendum:2011}.
All experimentally determined properties were reproduced theoretically
with an at least semi-quantitative accuracy. In particular, the applicability
range of the Kholodenko-Douglas generalized Stokes-Einstein (KD-GSE)
relation \cite{KholodenkoDouglas:1995},

\[
\frac{d_{C}\left(\phi\right)\;\!\eta\left(\phi\right)}{d_{0}\;\!\eta_{0}}\;\!\sqrt{S(q\to0,\phi)}\;\!=\;\!1\;\!,
\]
between collective diffusion coefficient, steady-state viscosity $\eta=\eta_{\infty}+\Delta\eta$
(with $\eta_{\infty}=\eta_{0}+\Delta\eta_{\infty})$, and the square-root
of the isothermal osmotic compressibility coefficient $S(q\to0,\phi)$
was studied. The KD-GSE relation has been used in the biophysics and
soft matter communities to deduce the viscosity from static and dynamic
light scattering (SLS and DLS) data \cite{Nettesheim2008,Gaigalas1995,Cohen1998}.
Note that the short-time (superscript S) and long-time (superscript
L) forms of $d_{C}$ are practically equal to each other, whereas
$\eta_{\infty}<\eta$.

Fig. \ref{fig:BSA-GSE} provides a theoretical-experimental performance
test of the short-time and long-time versions of the KD-GSE relation
for BSA solutions at low salinity, i.e. $n_{s}=1-3$ mM, and for the
physiological salt concentration $n_{s}=150$ mM. The experimental
data agree overall well with the theoretical predictions. For low
salinity and larger protein concentrations $c_{p}$, pronounced deviations
from the protein concentration independent value one of the KD-GSE
relation are observed. A pronounced violation of this relation is
predicted theoretically also for suspensions of large charge-stabilized
colloidal spheres. For neutral hard-sphere suspensions up to $\phi\lesssim0.4$,
the KD-GSE relation is, however, decently well fulfilled \cite{Heinen:2011if}.
An effective charged-sphere model analysis similar to that for BSA
was made in a joint experimental-theoretical study of suspensions
of charged gibbsite platelets (with aspect ratio 1 : 11) in the isotropic
liquid phase \cite{Holmqvist_Gibbsite:2012}. As shown in this work,
the effect of translation-rotation interparticle coupling (disregarded
in the effective sphere treatment) is less pronounced for collective
diffusion, but it is particularly strong for translational and rotational
self-diffusion at larger concentrations.

\section{Concentrated charge-stabilized colloidal suspensions \label{sec:Charge-stabilized-colloidal-suspensions}}

We commence this section by shortly summarizing recent progress made
in developing computer simulation and numerical schemes allowing to
quantify microion electrokinetic effects on transport properties of
charge-stabilized suspensions at non-zero concentrations. In the smoothed
profile method (SMP) of Yamamoto, Nakayama and Kim \cite{Kim:2006,Yamamoto:2009},
and in the related fluid particle dynamics (FPD) method of Tanaka
and Araki \cite{Araki_Tanaka:2008}, the solvent and the microions
are treated as continuous fields, like in the SET approach. Only the
colloidal macroions are treated explicitly as particles, i.e. as high-viscosity
liquid droplets in FPD, and as particles with a smoothed interface
to the solvent in SMP. The particles in both methods act on the solvent
through continuous body forces rather than through moving boundaries.
This simplification results in efficient numerical solution schemes
suitable for concentrated suspensions. The advantage of the SMP method
is that larger time increments can be selected. Since the microions
in both methods are described in a mean-field way, local charge ordering
effects beyond the Poisson-Boltzmann (PB) level are not considered.
Both methods have been applied in particular to colloidal electrophoresis.
The electrophoretic mobility results by the SMP method \cite{Kim:2006}
are similar to those obtained by SET spherical cell model calculations
where overlapping colloidal EDLs are approximately accounted for \cite{Carrique200543}.
Giupponi and Pagonabarraga solve the (non-linearized) SET equations
by a discretized lattice Boltzmann (LB) formulation of the solvent, coupled to the colloid
surface grid points by kinetic bounce-back rules, and combined with
a solver of the correspondingly discretized SET convection-diffusion
equations for the microion densities \cite{Giupponi:2011}. They could
show that for microion Peclet numbers $\gtrsim0.3$ where flow advection
becomes comparably important to single-microion diffusion, non-linear
microion advection enhances the electrophoretic mobility.

\begin{figure*}
\begin{center}
\resizebox{0.8\textwidth}{!}{
\includegraphics{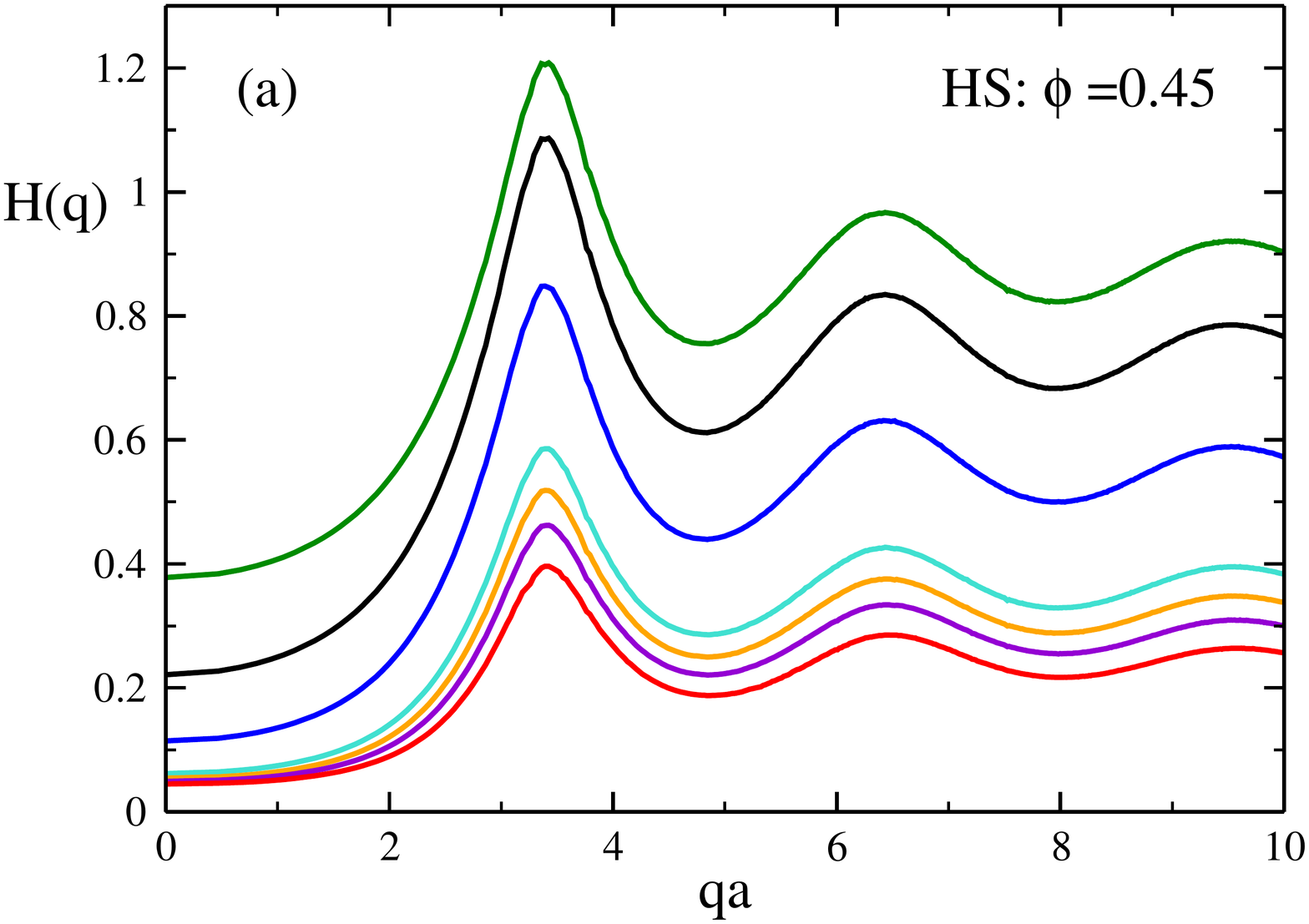}\includegraphics{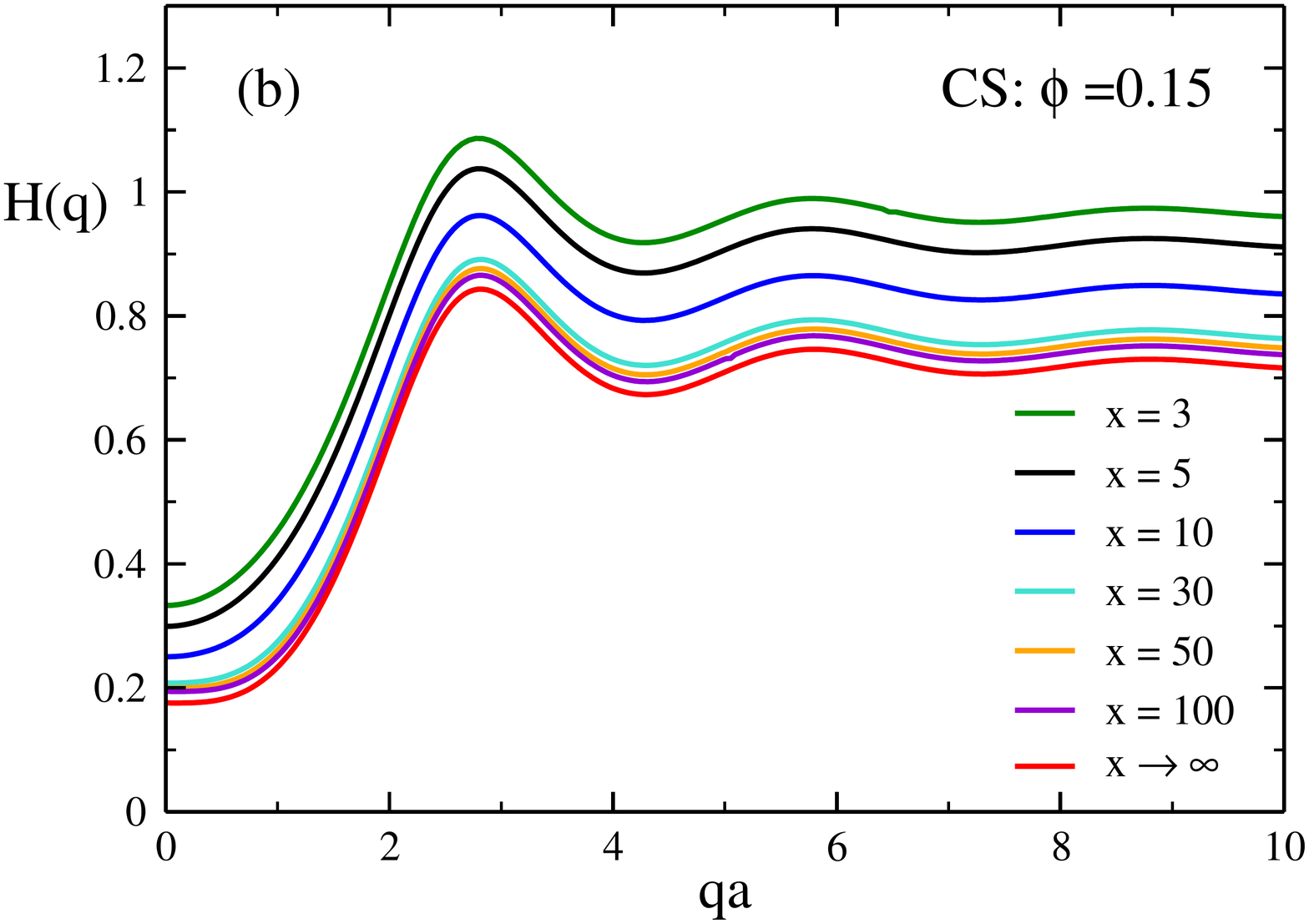}}
\end{center}
\vspace{-2em}
\caption{Wavenumber-dependent hydrodynamic function
$H(q)$, of solvent-permeable colloidal spheres, with values of the
inverse permeability coefficient $x$ and volume fraction $\phi$
as indicated. Small $x$ correspond to large, and large $x$ to small
permeabilities. (a) Simulation results for uncharged hard spheres
taken from \cite{Abade:2010gt}. The curve for stick hydrodynamic
BC ($x=\infty$) agrees with the ASD simulation data. (b) Analytic
results for charged spheres of diameter $2a=200$ nm and effective
particle charge number $Z=200$, immersed in a $0.5$ mM $1:1$ aqueous
electrolyte solution with Bjerum length $L_{B}=7.11$ nm. The charged
spheres interact by a screened Coulomb potential of DLVO type. The
results in (b) were obtained by a pairwise additive HIs approach,
with truncated pair mobilities of ${\cal O}\left(1/r^{7}\right)$
in the pair distance $r$ taken from \cite{1978PhyA...93..465R},
and using the accurate MPB-RMSA static structure factor input. }
\label{fig:Hq-permeable-HS-CS}
\end{figure*}

Lobaskin, D\"unweg and collaborators \cite{Lobaskin:2004,Lobaskin_PRL:2007}
have developed a hybrid simulation method where the microions are
considered explicitly, and where the raspberry-like macroion model
surface is coupled through a friction term to a LB background describing
the Navier-Stokes hydrodynamics of a structureless
solvent. While this simulation method includes microion correlations
beyond the PB level, the price to pay is a larger numerical effort.
Due to the more costly numerics, only the electrophoresis of a single
macroion plus its neutralizing microion cloud in a box with periodic
BCs has been considered so far. Non-zero colloid concentration effects
are accounted for, akin to standard cell model calculations, by adjusting
the box to macroion size ratio to the given volume fraction. A similar
Molecular Dynamics - LB hybrid simulation method was used by Chatterji
and Horbach to analyze the nature of the effective electrophoretic
macroion charge in single-colloid electrophoresis \cite{Chatterji:2011}.
To describe electrophoresis of a spherical macroion in a finite box
on level of the mean-field SET equations, D\"unweg \emph{et al}. have
developed a numerically efficient solver \cite{Duenweg:2012}, and
they have used in addition the finite-element software package COMSOL
\cite{Duenweg:2008}.

Except for electrophoresis and sedimentation, microionic electrokinetic
contributions to linear colloidal transport coefficients are secondary
effects in magnitude, in particular for suspensions of strongly correlated
charged colloidal particles. Using a simplified MCT-HIs method, this
has been shown quantitatively in \cite{McPhie:2007ee,McPhie:2004gy}
for the long-time colloidal self-diffusion coefficient. In the remainder
of this section, we discuss colloidal diffusion properties and the
viscosity of concentrated suspensions, which are well described on
basis of the OMF model of microion-dressed macroions. In this model,
colloidal dynamic properties can be obtained to high accuracy using
the accelerated Stokesian Dynamics (ASD) simulation method for charge-stabilized
Brownian spheres \cite{2005JChPh.123e4708G,Banchio:2008gt,Heinen_Appl_cryt:2010,Heinen:2011if}
which fully accounts for the colloid HIs including lubrication interactions.
It was used so far mainly for the calculation of short-time diffusion
properties of charge-stabilized colloids, giving results in quantitative
agreement with dynamic scattering data (see \cite{Heinen_Appl_cryt:2010,Westermeier:2012}
for an elaborate comparison). Recently, (A)SD results have been obtained
also for long-time diffusion transport coefficients, and for self-intermediate
and collective dynamic scattering functions \cite{Banchio_Scaling:2013}.

\begin{figure}
\begin{center}
\resizebox{1.0\columnwidth}{!}{
\includegraphics{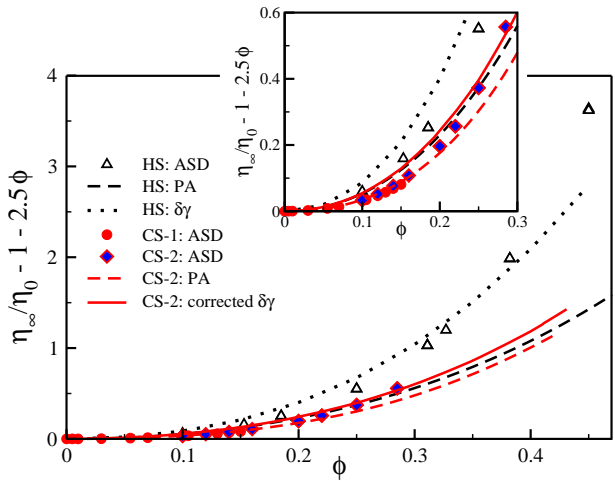}
}
\end{center}
\vspace{-1em}
\caption{Reduced high-frequency viscosity, $\eta_{\infty}/\eta_{0}$,
as function of $\phi$, for a neutral hard-sphere suspension (HS,
in black), and two deionized charged-sphere suspensions (CS-1 and
CS-2, in red). The leading-order Einstein viscosity contribution,
$1+2.5\phi$, is subtracted to expose the differences. Symbols: ASD
simulation results. Dashed lines: PA-approximation results. Dotted
lines: $\delta\gamma$-scheme results. Solid lines: self-part corrected
$\delta\gamma$-scheme results. All analytic schemes use the MPB-RMSA
$S(q)$ as input. The CS-1 results represented by red filled circles
are ASD data for $L_{B}=5.617$ nm, $\sigma=200$ nm, and $Z=100$.
The ASD data for the more weakly charged and smaller particles of
system CS-2, where $L_{B}=0.71$ nm, $\sigma=50$ nm, and $Z=70$,
are indicated by red diamonds filled in blue. The parameters of system
CS-2 have been used in the analytic calculations. The inset magnifies
details at lower $\phi$. Figure taken from \cite{Heinen:2011if}.}
\label{fig:eta-infty-CS-HS-1}
\end{figure}

We mention here in addition the powerful hydrodynamic force multipole
simulation method by Cichocki, Ekiel-Jezewska, Wajnryb and coworkers
\cite{Cichocki_HYDROMULTIPOLE:1994}, encoded in the HYDROMULTIPOLE
program, which allows for an easy implementation of different hydrodynamic
particle models. The method has been used for a comprehensive simulation
study of short-time dynamic properties of suspensions of solvent-permeable
hard spheres, with uniform permeability \cite{Abade:2010gt,Abade_JPCM_2010,AbadeViscosityJCP:2010},
and also with internal core-shell structure \cite{AbadeCoreshellJCP:2012}.
The flow inside the permeable region of a particle is described by
the Brinkman-Debye-Bueche fluid model. Results have been discussed
including the high-frequency viscosity $\eta_{\infty}$, and the hydrodynamic
function $H(q)$. The latter quantity reduces to the normalized mean
sedimentation velocity for small wavenumber $q$, and to the short-time
self-diffusion coefficient $d_{S}$ for large $q$.

HYDROMULTIPOLE results for the $H(q)$ of a concentrated suspension
of solvent-permeable, i.e. porous, hard spheres of excluded volume
radius $a$ are depicted in Fig. \ref{fig:Hq-permeable-HS-CS}(a),
for various values of the inverse permeability coefficient $x=a/l_{perm}.$
Here, $l_{perm}$ is the hydrodynamic penetration depth. As it is
noticed, (short-time) self- and collective diffusion, and sedimentation,
are significantly enhanced with increasing solvent-permeability of
the particles. The high-frequency viscosity, on the other hand, is
lowered. In \cite{Abade_JPCM_2010}, it is shown through comparison
with precise HYDROMULTIPOLE simulation data for $\eta_{\infty}$ that
the cell model approach for permeable and non-permeable spheres \cite{Ohshima:2009}
gives rather poor predictions for the concentration dependence of
$\eta_{\infty}$.

The force multipole method by Chichocki and collaborators has been
applied so far to electrically neutral solvent-permeable colloids
only. At lower $\phi$, however, non-pairwise additive HIs contributions
are small for charged particles due to their longer range mutual repulsion.
These contributions can thus be neglected to a decent approximation.
Fig. \ref{fig:Hq-permeable-HS-CS}(b) shows new hydrodynamic function
results for permeable charged spheres, which we have obtained using
truncated two-sphere hydrodynamic mobility tensors in combination
with the MPB-RMSA static structure factor input. According to the
figure, the influence of solvent permeability on $H(q)$ is substantially
weaker for charged colloidal particles.

In elaborate studies, it has been shown that the ASD simulation results
for short-time diffusion properties of charge-stabilized spheres with
stick hydrodynamic BC, and a large body of experimental data alike
\cite{Heinen_Appl_cryt:2010,Westermeier:2012}, are well reproduced,
for a broad range of interaction parameters and volume fractions typical
of the liquid-phase state, using the analytic and easy-to-implement
$\delta\gamma$ method by Beenakker and Mazur, amended by an improved
(i.e., ``corrected'') self-part contribution \cite{Heinen:2011if}.
Makuch and Cichocki \cite{Makuch:2012} have critically assessed the
approximations going into the Beenakker-Mazur method. Their revised
version of the $\delta\gamma$ method which includes an improved hydrodynamic
mobility matrix input leads overall to larger differences from the
hard-sphere simulation data. This points to a fortuitous cancellation
of errors introduced by the approximations going into the original
Beenakker-Mazur method.

A self-part corrected version of the $\delta\gamma$ method has been
successfully used in addition for the calculation of the high-frequency
viscosity of charge-stabilized colloids at lower salinity. Results
by this method are shown in Fig. \ref{fig:eta-infty-CS-HS-1}, and
compared with ASD simulation results for charge-stabilized spheres
(CS) and neutral hard spheres (HS) with stick hydrodynamic BC. Also
included are viscosity results obtained using numerically accurate
values for the full pairwise-additive two-sphere hydrodynamic mobilities
including lubrication forces. This is referred to as the PA-approximation.
While $\eta_{\infty}$is for CS smaller than for HS, the opposite
ordering is valid in general for the steady-state, low-shear rate
viscosity $\eta$, owing to the for CS substantially larger colloidal
shear-stress relaxation contribution $\Delta\eta$. Possible correlations
between $\eta$ and $\eta_{\infty}$, and the Debye screening length
and macroion electric surface potential, are discussed in \cite{RusselViscosity:2009}
on basis of theories for semidilute systems with and without single-colloid
electrokinetics included.

\begin{figure}
\begin{center}
\resizebox{1.0\columnwidth}{!}{
\includegraphics{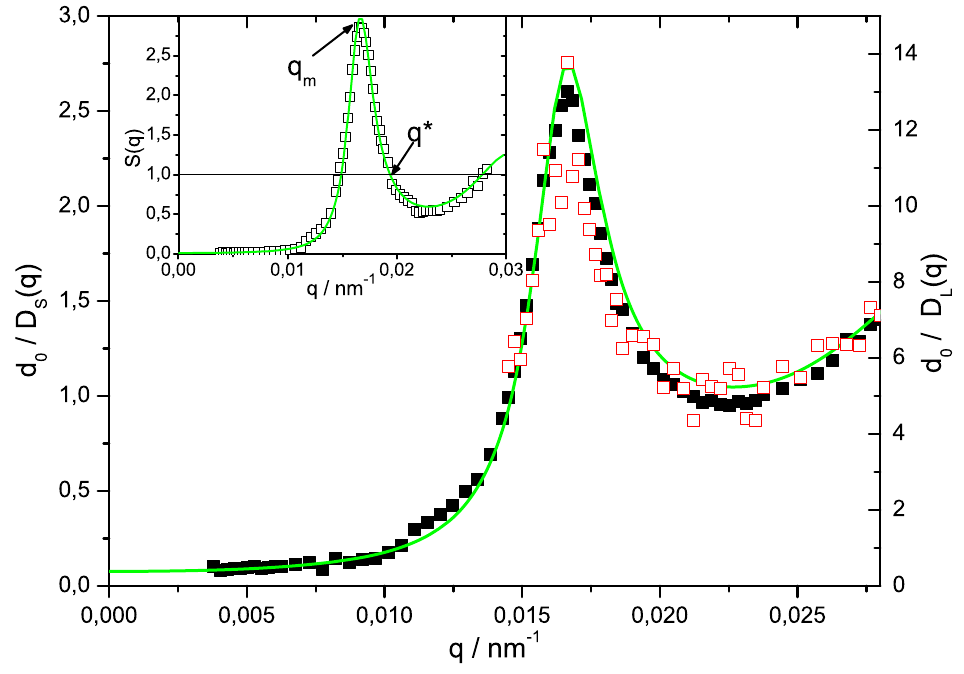}
}
\end{center}
\vspace{-1em}
\caption{Dynamic light scattering results for
the normalized inverse short-time diffusion function $d_{0}/D_{S}(q)$
(filled black squares, scale on left side) and inverse long-time diffusion
function $d_{0}/D_{L}(q)$ (open red squares, scale on right side)
of charge-stabilized silica spheres at $\phi=0.14$. Solid curve:
theoretical prediction by the $\delta\gamma$-scheme for the wavenumber
dependence of $d_{0}/D_{S}(q)$, using the MPB-RMSA input for $S(q)$
shown in the inset together with the experimental data. Figure taken
from \cite{Holmqvist_PRL:2010}.}
\label{fig:Dynamic-Scaling}
\end{figure}

A dynamic scaling relation for the normalized dynamic structure factor,

\[
S(q,t)/S(q)\;\!\approx\;\!\exp\left\{ -q^{2}\;\!\frac{D_{S}(q)}{d_{S}}\;\! W(t)\right\} \,\,,
\]
has been shown in \cite{Holmqvist_PRL:2010} to apply approximately
to charge-stabilized colloids, within the experimentally accessed
correlation time window, and for scattering wavenumbers $q$ around
the position, $q_{m}$, of the principal static structure factor peak.
Here, $W(t)=\left\langle \Delta r^{2}\left(t\right)\right\rangle /6$
is the particle mean-squared displacement with short-time slope $d_{S}$,
and $D_{S}(q)=d_{0}\;\! H(q)/S(q)$ is the short-time diffusion function.
The scaling relation was initially found empirically by Segr\`{e} and
Pusey for the case of colloidal hard spheres \cite{SegrePusey_PRL:1996} (see here also
the MCT study of Fuchs and Mayr \cite{Fuchs1999}).
It implies in particular that $D_{L}(q)/D_{S}(q)\approx d_{L}/d_{S}$,
for values of $q$ near to $q_{m}$. Here, $D_{L}(q)$ denotes the
relaxation rate of the long-time decay of $S(q,t)$ which in the experimental
time window is single-exponential, and $d_{L}$ is the colloidal long-time
self-diffusion coefficient which is smaller than $d_{S}$. Fig. \ref{fig:Dynamic-Scaling}
demonstrates that this relation is indeed valid to a decent approximation
for charge-stabilized spheres. The validity of the time-wavenumber
factorization scaling of $S(q,t)$ has been repeatedly challenged,
also for the reason that the true long-time regime can hardly be reached
experimentally \cite{Lurio_PRL:2000,MartinezScaling:2011}). As it
was argued already in \cite{Holmqvist_PRL:2010}, from a theoretical
viewpoint dynamic scaling is merely an approximate feature. This viewpoint
is corroborated by a recent study, where Stokesian dynamics, Brownian
dynamics, and MCT calculations of $S(q,t), W(t)$ and $D_{S}(q)$ have been compared \cite{Banchio_Scaling:2013}.

\begin{figure}
\begin{center}
\resizebox{1.0\columnwidth}{!}{
\includegraphics{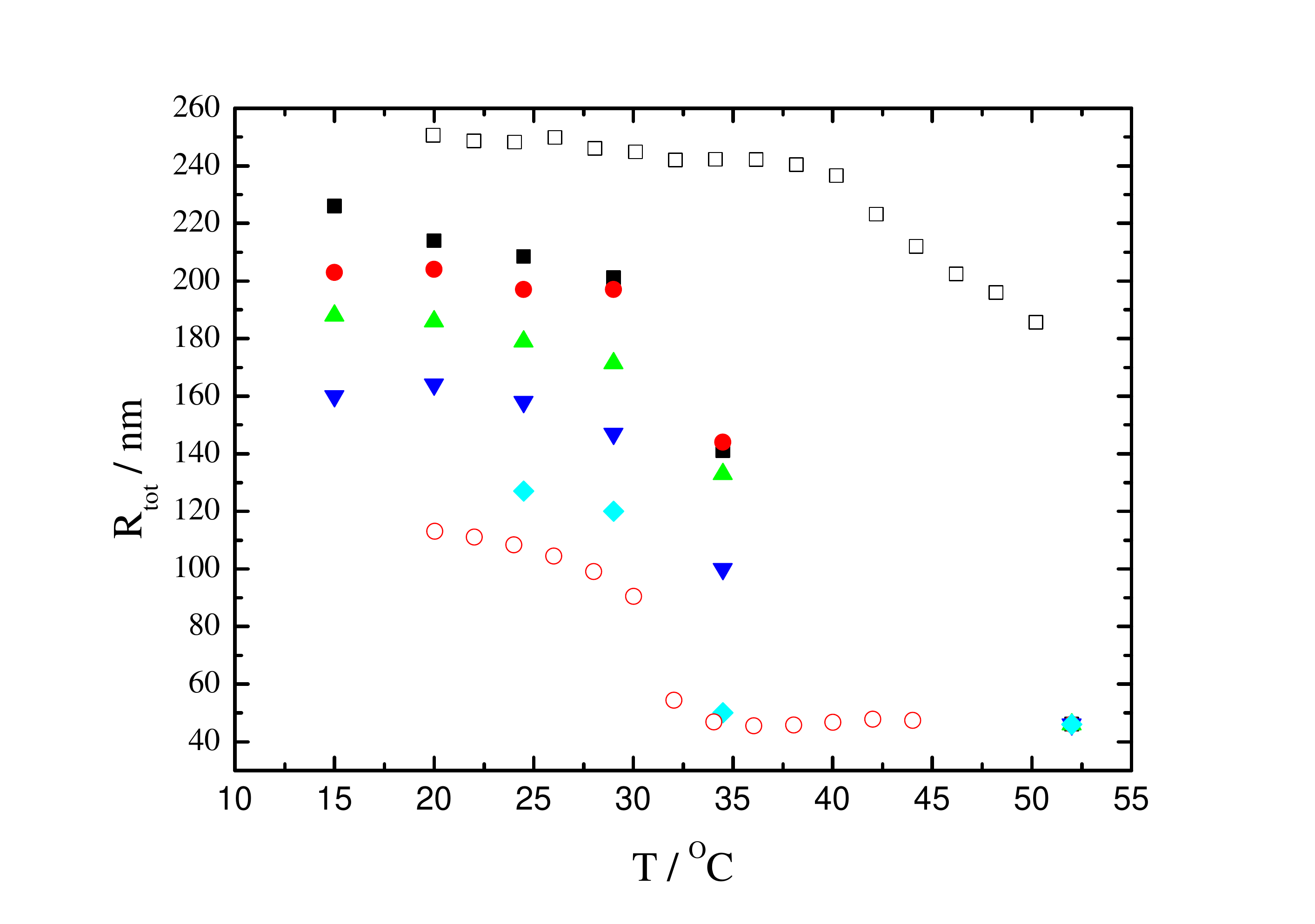}
}
\end{center}
\vspace{-1em}
\caption{Outer radius, $R_{tot}$, as a function
of $T$ for ionic PNiPAm dispersions (filled symbols) at microgel
particle concentration $n_{p}=0.0053$ (black squares), $0.0074$
(red circles), $0.013$ (green upper triangles), $0.022$ (blue lower
triangles) and $0.069$ M (light blue diamonds). Open symbols: Results for a dilute particle solutions
with $n_{p}=0.1$ nM and pH = 6.3 (black), and pH = 3.0 (red). Figure
taken from \cite{Holmqvist_PRL:2012}.}
\label{fig:Microgel-Radius}
\end{figure}

Ionic microgels belong to the subclass of ultra-soft colloids where
the effective interaction potential crosses over from a screened Coulomb
repulsion at non-overlap distances to a smoothly increasing repulsion
of finite interaction energy for overlapping particles \cite{Denton:2003,Likos_MicrogelReview:2011}.
A fraction of the counterions released by ionic polymer backbone groups
of the microgel network is confined to the interior of the ionic microgel
particles, affecting therefore their swelling behavior through the
microion osmotic pressure difference. A comprehensive experimental-theoretical
study of the concentration- and temperature dependent swelling behavior
of PNiPAm ionic core-shell microgel particles was made in \cite{Holmqvist_PRL:2012}.
Results by this study for the outer microgel radius $R_{tot}$, deduced
as a function of microgel number density $n_{p}$ and temperature
$T$, are included in Fig. \ref{fig:Microgel-Radius}. In the theoretical
part of the study, state-of-the art analytic OMF methods discussed
also in relation to Figs. \ref{fig:BSA-GSE} - \ref{fig:Dynamic-Scaling}
have been used. Fig. \ref{fig:Microgel-Radius} highlights in particular
the strong concentration dependence of the (outer) microgel radius
in the non-collapsed microgel state at lower temperatures.

\section{Conclusions \label{sec:Conclusions}}

We have reviewed recent advances in our understanding of linear diffusion-convection
and viscoelastic transport properties of dispersions of charged Brownian
particles. Various theoretical and computer simulation methods have
been discussed and compared. Regarding the electrokinetics of electrolyte
solutions, a fast and versatile simplified MCT-HIs method is now available,
making predictions in good agreement with experimental data. With
some effort, this method can be extended to frequency-dependent transport
properties, and to size- and charge-asymmetric electrolytes. Work
on these extensions is in progress. The MCT-HIs method keeps promise
to tackle also the electrophoresis and conductivity of colloidal macroions
and globular proteins with overlapping EDLs. The microions are hereby
treated on equal footing with the macroions, both statically and dynamically.
While great progress has been made in the development of numerical
cell model and simulation schemes describing colloidal electrokinetics,
the proper inclusion of electro-steric and hydrodynamic aspects of
the microion dynamics beyond the mean-field level is still a challenging
task, in particular for concentrated suspensions. A detailed evaluation
of the pros and cons of the various existing simulation schemes for
selected colloidal electrophoresis benchmark problems is still on
demand. Short-time dynamic properties treated on the level of the
OMF model are meanwhile well understood, thanks to detailed experimental
and (A)SD simulation results. Well-tested accurate theoretical methods
for calculating short-time dynamic properties such as the self-part
corrected $\delta\gamma$ scheme are now available. They allow for
a fast analysis of experimental data covering a broad range of system
parameters. Short-time dynamic properties are of relevance not only
in their own right but are required also as input to theories describing
colloidal long-time dynamics in concentrated systems such as MCT and
DDFT approaches, and their hybrids. A lot remains to be learned about
the long-time dynamics of charge-stabilized suspensions and globular
protein solutions, in particular when macroions with internal electro-steric
and hydrodynamic structure are considered.

\subsection*{acknowledgement}
We thank G. Abade, B. Cichocki, J. Dhont, S. Egelhaaf, M. Ekiel-Jezewska,
J. Gapinksi, Ch. G\"ogelein, P. Holmqvist, K. Makuch, M. McPhie, P.S.
Mohanty, T. Palberg, A. Patkowski, J. Riest, R. Roa, F. Schreiber,
P. Schurtenberger, and E. Wajnryb for helpful discussions and fruitful
collaborations.
Financial support by the Deutsche Forschungsgemeinschaft (SFB-TR6,
Project B2) is gratefully acknowledged.

\subsection*{Abbreviations:}
\renewcommand{\arraystretch}{1.25}
\begin{longtable}{p{.25\columnwidth}p{.75\columnwidth}}
ASD \dotfill&Accelerated Stokesian Dynamics.\\
BSA \dotfill& Bovine serum albumin.\\
BCs \dotfill& Boundary conditions.\\
CS \dotfill& Charged spheres.\\
DDFT \dotfill& Dynamic density functional theory.\\
DFOF \dotfill& Debye-Falkenhagen-Onsager-Fuoss.\\
DH \dotfill& Debye-H\"{u}ckel.\\
DLS \dotfill& Dynamic light scattering.\\
DLVO \dotfill& Derjaguin-Landau-Verwey-Overbeek.\\
EDLs \dotfill& Electric double layers.\\
FPD \dotfill& Fluid particle dynamics.\\
GSmE \dotfill& Generalized Smoluchowski equation.\\
HIs \dotfill& Hydrodynamic interactions.\\
HNC \dotfill& Hypernetted chain.\\
HS \dotfill& Uncharged, pairwise additive hard spheres.\\
KD-GSE \dotfill& Kholodenko-Douglas generalized Stokes-Einstein (relation).\\
LB \dotfill& Lattice-Boltzmann.\\
MCT \dotfill& Mode-coupling theory.\\
MPB-RMSA \dotfill& Modified penetrating background-corrected rescaled mean spherical approximation.\\
MSA \dotfill& Mean spherical approximation.\\
NE \dotfill& Nernst-Einstein.\\
OMF \dotfill& One-component macroion fluid.\\
PB \dotfill& Poisson-Boltzmann.\\
PM \dotfill& Primitive model.\\
PNiPAm \dotfill& Poly (N-isopropylacrylamide).\\
SET \dotfill& Standard electrokinetic theory.\\
SLS \dotfill& Static light scattering.\\
SMP \dotfill& Smoothed profile method.\\
\end{longtable}
~\newline

\bibliography{EPJE_Naegele_2013}

\begin{thebibliography}{117}%
\makeatletter
\providecommand \@ifxundefined [1]{%
 \@ifx{#1\undefined}
}%
\providecommand \@ifnum [1]{%
 \ifnum #1\expandafter \@firstoftwo
 \else \expandafter \@secondoftwo
 \fi
}%
\providecommand \@ifx [1]{%
 \ifx #1\expandafter \@firstoftwo
 \else \expandafter \@secondoftwo
 \fi
}%
\providecommand \natexlab [1]{#1}%
\providecommand \enquote  [1]{``#1''}%
\providecommand \bibnamefont  [1]{#1}%
\providecommand \bibfnamefont [1]{#1}%
\providecommand \citenamefont [1]{#1}%
\providecommand \href@noop [0]{\@secondoftwo}%
\providecommand \href [0]{\begingroup \@sanitize@url \@href}%
\providecommand \@href[1]{\@@startlink{#1}\@@href}%
\providecommand \@@href[1]{\endgroup#1\@@endlink}%
\providecommand \@sanitize@url [0]{\catcode `\\12\catcode `\$12\catcode
  `\&12\catcode `\#12\catcode `\^12\catcode `\_12\catcode `\%12\relax}%
\providecommand \@@startlink[1]{}%
\providecommand \@@endlink[0]{}%
\providecommand \url  [0]{\begingroup\@sanitize@url \@url }%
\providecommand \@url [1]{\endgroup\@href {#1}{\urlprefix }}%
\providecommand \urlprefix  [0]{URL }%
\providecommand \Eprint [0]{\href }%
\providecommand \doibase [0]{http://dx.doi.org/}%
\providecommand \selectlanguage [0]{\@gobble}%
\providecommand \bibinfo  [0]{\@secondoftwo}%
\providecommand \bibfield  [0]{\@secondoftwo}%
\providecommand \translation [1]{[#1]}%
\providecommand \BibitemOpen [0]{}%
\providecommand \bibitemStop [0]{}%
\providecommand \bibitemNoStop [0]{.\EOS\space}%
\providecommand \EOS [0]{\spacefactor3000\relax}%
\providecommand \BibitemShut  [1]{\csname bibitem#1\endcsname}%
\let\auto@bib@innerbib\@empty
\bibitem [{\citenamefont {Gapinski}\ \emph {et~al.}(2005)\citenamefont
  {Gapinski}, \citenamefont {Wilk}, \citenamefont {Patkowski}, \citenamefont
  {H{\"a}u{\ss}ler}, \citenamefont {Banchio}, \citenamefont {Pecora},\ and\
  \citenamefont {N{\"a}gele}}]{2005JChPh.123e4708G}%
  \BibitemOpen
  \bibfield  {author} {\bibinfo {author} {\bibfnamefont {J.}~\bibnamefont
  {Gapinski}}, \bibinfo {author} {\bibfnamefont {A.}~\bibnamefont {Wilk}},
  \bibinfo {author} {\bibfnamefont {A.}~\bibnamefont {Patkowski}}, \bibinfo
  {author} {\bibfnamefont {W.}~\bibnamefont {H{\"a}u{\ss}ler}}, \bibinfo
  {author} {\bibfnamefont {A.~J.}\ \bibnamefont {Banchio}}, \bibinfo {author}
  {\bibfnamefont {R.}~\bibnamefont {Pecora}}, \ and\ \bibinfo {author}
  {\bibfnamefont {G.}~\bibnamefont {N{\"a}gele}},\ }\href {\doibase
  10.1063/1.1996569} {\bibfield  {journal} {\bibinfo  {journal} {J. Chem.
  Phys.}\ }\textbf {\bibinfo {volume} {123}},\ \bibinfo {pages} {054708}
  (\bibinfo {year} {2005})}\BibitemShut {NoStop}%
\bibitem [{\citenamefont {Heinen}\ \emph {et~al.}(2012)\citenamefont {Heinen},
  \citenamefont {Zanini}, \citenamefont {Roosen-Runge}, \citenamefont
  {Fedunova}, \citenamefont {Zhang}, \citenamefont {Hennig}, \citenamefont
  {Seydel}, \citenamefont {Schweins}, \citenamefont {Sztucki}, \citenamefont
  {Antalik}, \citenamefont {Schreiber},\ and\ \citenamefont
  {N{\"a}gele}}]{2012SMat....8.1404H}%
  \BibitemOpen
  \bibfield  {author} {\bibinfo {author} {\bibfnamefont {M.}~\bibnamefont
  {Heinen}}, \bibinfo {author} {\bibfnamefont {F.}~\bibnamefont {Zanini}},
  \bibinfo {author} {\bibfnamefont {F.}~\bibnamefont {Roosen-Runge}}, \bibinfo
  {author} {\bibfnamefont {D.}~\bibnamefont {Fedunova}}, \bibinfo {author}
  {\bibfnamefont {F.}~\bibnamefont {Zhang}}, \bibinfo {author} {\bibfnamefont
  {M.}~\bibnamefont {Hennig}}, \bibinfo {author} {\bibfnamefont
  {T.}~\bibnamefont {Seydel}}, \bibinfo {author} {\bibfnamefont
  {R.}~\bibnamefont {Schweins}}, \bibinfo {author} {\bibfnamefont
  {M.}~\bibnamefont {Sztucki}}, \bibinfo {author} {\bibfnamefont
  {M.}~\bibnamefont {Antalik}}, \bibinfo {author} {\bibfnamefont
  {F.}~\bibnamefont {Schreiber}}, \ and\ \bibinfo {author} {\bibfnamefont
  {G.}~\bibnamefont {N{\"a}gele}},\ }\href {\doibase 10.1039/c1sm06242e}
  {\bibfield  {journal} {\bibinfo  {journal} {Soft Matter}\ }\textbf {\bibinfo
  {volume} {8}},\ \bibinfo {pages} {1404} (\bibinfo {year} {2012})}\BibitemShut
  {NoStop}%
\bibitem [{\citenamefont {Roosen-Runge}\ \emph {et~al.}(2011)\citenamefont
  {Roosen-Runge}, \citenamefont {Hennig}, \citenamefont {Zhang}, \citenamefont
  {Jacobs}, \citenamefont {Sztucki}, \citenamefont {Schober}, \citenamefont
  {Seydel},\ and\ \citenamefont {Schreiber}}]{RoosenRunge:2011}%
  \BibitemOpen
  \bibfield  {author} {\bibinfo {author} {\bibfnamefont {F.}~\bibnamefont
  {Roosen-Runge}}, \bibinfo {author} {\bibfnamefont {M.}~\bibnamefont
  {Hennig}}, \bibinfo {author} {\bibfnamefont {F.}~\bibnamefont {Zhang}},
  \bibinfo {author} {\bibfnamefont {R.~M.~J.}\ \bibnamefont {Jacobs}}, \bibinfo
  {author} {\bibfnamefont {M.}~\bibnamefont {Sztucki}}, \bibinfo {author}
  {\bibfnamefont {H.}~\bibnamefont {Schober}}, \bibinfo {author} {\bibfnamefont
  {T.}~\bibnamefont {Seydel}}, \ and\ \bibinfo {author} {\bibfnamefont
  {F.}~\bibnamefont {Schreiber}},\ }\href {\doibase 10.1073/pnas.1107287108}
  {\bibfield  {journal} {\bibinfo  {journal} {Proc. Natl. Acad. Sci. U.S.A.}\
  }\textbf {\bibinfo {volume} {108}},\ \bibinfo {pages} {11815} (\bibinfo
  {year} {2011})}\BibitemShut {NoStop}%
\bibitem [{\citenamefont {G{\"ogelein}}\ \emph {et~al.}(2008)\citenamefont
  {G{\"ogelein}}, \citenamefont {N{\"a}gele}, \citenamefont {Tuinier},
  \citenamefont {Gibaud}, \citenamefont {Stradner},\ and\ \citenamefont
  {Schurtenberger}}]{Goegelein:2008}%
  \BibitemOpen
  \bibfield  {author} {\bibinfo {author} {\bibfnamefont {C.}~\bibnamefont
  {G{\"ogelein}}}, \bibinfo {author} {\bibfnamefont {G.}~\bibnamefont
  {N{\"a}gele}}, \bibinfo {author} {\bibfnamefont {R.}~\bibnamefont {Tuinier}},
  \bibinfo {author} {\bibfnamefont {T.}~\bibnamefont {Gibaud}}, \bibinfo
  {author} {\bibfnamefont {A.}~\bibnamefont {Stradner}}, \ and\ \bibinfo
  {author} {\bibfnamefont {P.}~\bibnamefont {Schurtenberger}},\ }\href
  {\doibase 10.1063/1.2951987} {\bibfield  {journal} {\bibinfo  {journal} {J.
  Chem. Phys.}\ }\textbf {\bibinfo {volume} {129}},\ \bibinfo {pages} {085102}
  (\bibinfo {year} {2008})}\BibitemShut {NoStop}%
\bibitem [{\citenamefont {Egelhaaf}\ \emph {et~al.}(2004)\citenamefont
  {Egelhaaf}, \citenamefont {Lobaskin}, \citenamefont {Bauer}, \citenamefont
  {Merkle},\ and\ \citenamefont {Schurtenberger}}]{Egelhaaf2004}%
  \BibitemOpen
  \bibfield  {author} {\bibinfo {author} {\bibfnamefont {S.~U.}\ \bibnamefont
  {Egelhaaf}}, \bibinfo {author} {\bibfnamefont {V.}~\bibnamefont {Lobaskin}},
  \bibinfo {author} {\bibfnamefont {H.~H.}\ \bibnamefont {Bauer}}, \bibinfo
  {author} {\bibfnamefont {H.~P.}\ \bibnamefont {Merkle}}, \ and\ \bibinfo
  {author} {\bibfnamefont {P.}~\bibnamefont {Schurtenberger}},\ }\href
  {\doibase 10.1140/epje/e2004-00051-2} {\bibfield  {journal} {\bibinfo
  {journal} {Eur. Phys. J. E}\ }\textbf {\bibinfo {volume} {13}},\ \bibinfo
  {pages} {153} (\bibinfo {year} {2004})}\BibitemShut {NoStop}%
\bibitem [{\citenamefont {G{\"ogelein}}\ \emph {et~al.}(2012)\citenamefont
  {G{\"ogelein}}, \citenamefont {Wagner}, \citenamefont {Cardinaux},
  \citenamefont {N{\"a}gele},\ and\ \citenamefont {Egelhaaf}}]{Goegelein:2012}%
  \BibitemOpen
  \bibfield  {author} {\bibinfo {author} {\bibfnamefont {C.}~\bibnamefont
  {G{\"ogelein}}}, \bibinfo {author} {\bibfnamefont {D.}~\bibnamefont
  {Wagner}}, \bibinfo {author} {\bibfnamefont {F.}~\bibnamefont {Cardinaux}},
  \bibinfo {author} {\bibfnamefont {G.}~\bibnamefont {N{\"a}gele}}, \ and\
  \bibinfo {author} {\bibfnamefont {S.~U.}\ \bibnamefont {Egelhaaf}},\ }\href
  {\doibase 10.1063/1.3673442} {\bibfield  {journal} {\bibinfo  {journal} {J.
  Chem. Phys.}\ }\textbf {\bibinfo {volume} {136}},\ \bibinfo {pages} {015102}
  (\bibinfo {year} {2012})}\BibitemShut {NoStop}%
\bibitem [{\citenamefont {Davis}\ \emph {et~al.}(1991)\citenamefont {Davis},
  \citenamefont {Madura}, \citenamefont {Sines}, \citenamefont {Luty},
  \citenamefont {Allison},\ and\ \citenamefont
  {Mccamon}}]{ISI:A1991GV01300022}%
  \BibitemOpen
  \bibfield  {author} {\bibinfo {author} {\bibfnamefont {M.~E.}\ \bibnamefont
  {Davis}}, \bibinfo {author} {\bibfnamefont {J.~D.}\ \bibnamefont {Madura}},
  \bibinfo {author} {\bibfnamefont {J.}~\bibnamefont {Sines}}, \bibinfo
  {author} {\bibfnamefont {B.~A.}\ \bibnamefont {Luty}}, \bibinfo {author}
  {\bibfnamefont {S.~A.}\ \bibnamefont {Allison}}, \ and\ \bibinfo {author}
  {\bibfnamefont {J.~A.}\ \bibnamefont {Mccamon}},\ }\href {\doibase
  10.1016/0076-6879(91)02024-4} {\bibfield  {journal} {\bibinfo  {journal}
  {Methods in Enzymology}\ }\textbf {\bibinfo {volume} {202}},\ \bibinfo
  {pages} {473} (\bibinfo {year} {1991})}\BibitemShut {NoStop}%
\bibitem [{\citenamefont {Nonner}\ \emph {et~al.}(1999)\citenamefont {Nonner},
  \citenamefont {Chen},\ and\ \citenamefont {Eisenberg}}]{Nonner:1999hx}%
  \BibitemOpen
  \bibfield  {author} {\bibinfo {author} {\bibfnamefont {W.}~\bibnamefont
  {Nonner}}, \bibinfo {author} {\bibfnamefont {D.~P.}\ \bibnamefont {Chen}}, \
  and\ \bibinfo {author} {\bibfnamefont {B.}~\bibnamefont {Eisenberg}},\ }\href
  {\doibase 10.1085/jgp.113.6.773} {\bibfield  {journal} {\bibinfo  {journal}
  {J. Gen. Physiol.}\ }\textbf {\bibinfo {volume} {113}},\ \bibinfo {pages}
  {773} (\bibinfo {year} {1999})}\BibitemShut {NoStop}%
\bibitem [{\citenamefont {Jenkins}\ and\ \citenamefont
  {Marcus}(1995)}]{Jenkins:1995}%
  \BibitemOpen
  \bibfield  {author} {\bibinfo {author} {\bibfnamefont {H.~D.~B.}\
  \bibnamefont {Jenkins}}\ and\ \bibinfo {author} {\bibfnamefont
  {Y.}~\bibnamefont {Marcus}},\ }\href {\doibase 10.1021/cr00040a004}
  {\bibfield  {journal} {\bibinfo  {journal} {Chem. Rev.}\ }\textbf {\bibinfo
  {volume} {95}},\ \bibinfo {pages} {2695} (\bibinfo {year}
  {1995})}\BibitemShut {NoStop}%
\bibitem [{\citenamefont {Wolynes}(1980)}]{WOLYNES:1980}%
  \BibitemOpen
  \bibfield  {author} {\bibinfo {author} {\bibfnamefont {P.~G.}\ \bibnamefont
  {Wolynes}},\ }\href {\doibase 10.1146/annurev.pc.31.100180.002021} {\bibfield
   {journal} {\bibinfo  {journal} {Annu. Rev. Phys. Chem.}\ }\textbf {\bibinfo
  {volume} {31}},\ \bibinfo {pages} {345} (\bibinfo {year} {1980})}\BibitemShut
  {NoStop}%
\bibitem [{\citenamefont {Barthel}\ \emph {et~al.}(1998)\citenamefont
  {Barthel}, \citenamefont {Krienke},\ and\ \citenamefont
  {Kunz}}]{Barthel1998}%
  \BibitemOpen
  \bibfield  {author} {\bibinfo {author} {\bibfnamefont {J.~M.~G.}\
  \bibnamefont {Barthel}}, \bibinfo {author} {\bibfnamefont {H.}~\bibnamefont
  {Krienke}}, \ and\ \bibinfo {author} {\bibfnamefont {W.}~\bibnamefont
  {Kunz}},\ }\href@noop {} {\emph {\bibinfo {title} {{Physical Chemistry of
  Electrolyte Solutions}}}},\ \bibinfo {series} {Topics in Physical Chemistry},
  Vol.~\bibinfo {volume} {5}\ (\bibinfo  {publisher} {Steinkopff},\ \bibinfo
  {address} {Darmstadt},\ \bibinfo {year} {1998})\BibitemShut {NoStop}%
\bibitem [{\citenamefont {Falkenhagen}\ and\ \citenamefont
  {Ebeling}(1971)}]{Falkenhagen:1971vq}%
  \BibitemOpen
  \bibfield  {author} {\bibinfo {author} {\bibfnamefont {H.}~\bibnamefont
  {Falkenhagen}}\ and\ \bibinfo {author} {\bibfnamefont {W.}~\bibnamefont
  {Ebeling}},\ }\href@noop {} {\enquote {\bibinfo {title} {{Theorie der
  Elektrolyte}},}\ }\bibinfo {howpublished} {S. Hirzel Verlag},\ \bibinfo
  {address} {Stuttgart} (\bibinfo {year} {1971})\BibitemShut {NoStop}%
\bibitem [{\citenamefont {N{\"a}gele}(2013)}]{NaegeleVarena2013}%
  \BibitemOpen
  \bibfield  {author} {\bibinfo {author} {\bibfnamefont {G.}~\bibnamefont
  {N{\"a}gele}},\ }in\ \href@noop {} {\emph {\bibinfo {booktitle} {In: Physics
  of Complex Colloids - Proceedings of International School of Physics Enrico
  Fermi}}},\ \bibinfo {editor} {edited by\ \bibinfo {editor} {\bibfnamefont
  {C.}~\bibnamefont {Bechinger}}, \bibinfo {editor} {\bibfnamefont
  {F.}~\bibnamefont {Sciortino}}, \ and\ \bibinfo {editor} {\bibfnamefont
  {P.}~\bibnamefont {Ziherl}}}\ (\bibinfo  {publisher} {IOS Amsterdam; SIF
  Bologna},\ \bibinfo {year} {2013})\BibitemShut {NoStop}%
\bibitem [{\citenamefont {Ohshima}(2006)}]{Ohshima:2006tk}%
  \BibitemOpen
  \bibfield  {author} {\bibinfo {author} {\bibfnamefont {H.}~\bibnamefont
  {Ohshima}},\ }\href@noop {} {\emph {\bibinfo {title} {{Theory of Colloid and
  Interfacial Electric Phenomena}}}},\ \bibinfo {series} {Interface Science and
  Technology}, Vol.~\bibinfo {volume} {12}\ (\bibinfo  {publisher} {Elsevier
  Academic Press},\ \bibinfo {address} {Amsterdam},\ \bibinfo {year}
  {2006})\BibitemShut {NoStop}%
\bibitem [{\citenamefont {Masliyah}\ and\ \citenamefont
  {Bhattacharjee}(2006)}]{BookMaslijay:2006}%
  \BibitemOpen
  \bibfield  {author} {\bibinfo {author} {\bibfnamefont {J.~H.}\ \bibnamefont
  {Masliyah}}\ and\ \bibinfo {author} {\bibfnamefont {S.}~\bibnamefont
  {Bhattacharjee}},\ }\href@noop {} {\emph {\bibinfo {title} {Electrokinetic
  and Colloid Transport Phenomena}}}\ (\bibinfo  {publisher} {Johns Wiley \&
  Sons, Hoboken, New Jersey},\ \bibinfo {year} {2006})\BibitemShut {NoStop}%
\bibitem [{\citenamefont {Bazant}\ \emph {et~al.}(2009)\citenamefont {Bazant},
  \citenamefont {Kilic}, \citenamefont {Storey},\ and\ \citenamefont
  {Ajdari}}]{Bazant200948}%
  \BibitemOpen
  \bibfield  {author} {\bibinfo {author} {\bibfnamefont {M.~Z.}\ \bibnamefont
  {Bazant}}, \bibinfo {author} {\bibfnamefont {M.~S.}\ \bibnamefont {Kilic}},
  \bibinfo {author} {\bibfnamefont {B.~D.}\ \bibnamefont {Storey}}, \ and\
  \bibinfo {author} {\bibfnamefont {A.}~\bibnamefont {Ajdari}},\ }\href
  {\doibase 10.1016/j.cis.2009.10.001} {\bibfield  {journal} {\bibinfo
  {journal} {Adv. Colloid Interface Sci.}\ }\textbf {\bibinfo {volume} {152}},\
  \bibinfo {pages} {48} (\bibinfo {year} {2009})}\BibitemShut {NoStop}%
\bibitem [{\citenamefont {Khair}\ and\ \citenamefont
  {Squires}(2009)}]{FLM:6774480}%
  \BibitemOpen
  \bibfield  {author} {\bibinfo {author} {\bibfnamefont {A.~S.}\ \bibnamefont
  {Khair}}\ and\ \bibinfo {author} {\bibfnamefont {T.~M.}\ \bibnamefont
  {Squires}},\ }\href {\doibase 10.1017/S0022112009991728} {\bibfield
  {journal} {\bibinfo  {journal} {J. Fluid. Mech.}\ }\textbf {\bibinfo {volume}
  {640}},\ \bibinfo {pages} {343} (\bibinfo {year} {2009})}\BibitemShut
  {NoStop}%
\bibitem [{\citenamefont {Roa}\ \emph {et~al.}(2011)\citenamefont {Roa},
  \citenamefont {Carrique},\ and\ \citenamefont {Ruiz-Reina}}]{C1CP22359C}%
  \BibitemOpen
  \bibfield  {author} {\bibinfo {author} {\bibfnamefont {R.}~\bibnamefont
  {Roa}}, \bibinfo {author} {\bibfnamefont {F.}~\bibnamefont {Carrique}}, \
  and\ \bibinfo {author} {\bibfnamefont {E.}~\bibnamefont {Ruiz-Reina}},\
  }\href {\doibase 10.1039/c1cp22359c} {\bibfield  {journal} {\bibinfo
  {journal} {Phys. Chem. Chem. Phys.}\ }\textbf {\bibinfo {volume} {13}},\
  \bibinfo {pages} {19437} (\bibinfo {year} {2011})}\BibitemShut {NoStop}%
\bibitem [{\citenamefont {Lopez-Garcia}\ \emph {et~al.}(2011)\citenamefont
  {Lopez-Garcia}, \citenamefont {Aranda-Rascon}, \citenamefont {Grosse},\ and\
  \citenamefont {Horno}}]{LopezGarcia:2011}%
  \BibitemOpen
  \bibfield  {author} {\bibinfo {author} {\bibfnamefont {J.~J.}\ \bibnamefont
  {Lopez-Garcia}}, \bibinfo {author} {\bibfnamefont {M.~J.}\ \bibnamefont
  {Aranda-Rascon}}, \bibinfo {author} {\bibfnamefont {C.}~\bibnamefont
  {Grosse}}, \ and\ \bibinfo {author} {\bibfnamefont {J.}~\bibnamefont
  {Horno}},\ }\href {\doibase 10.1016/j.jcis.2010.12.063} {\bibfield  {journal}
  {\bibinfo  {journal} {J. Colloid Interface Sci.}\ }\textbf {\bibinfo {volume}
  {356}},\ \bibinfo {pages} {325} (\bibinfo {year} {2011})}\BibitemShut
  {NoStop}%
\bibitem [{\citenamefont {Lozada-Cassou}\ \emph {et~al.}(1999)\citenamefont
  {Lozada-Cassou}, \citenamefont {Gonzales-Tovar},\ and\ \citenamefont
  {Olivares}}]{LozadaCassou:1999}%
  \BibitemOpen
  \bibfield  {author} {\bibinfo {author} {\bibfnamefont {M.}~\bibnamefont
  {Lozada-Cassou}}, \bibinfo {author} {\bibfnamefont {E.}~\bibnamefont
  {Gonzales-Tovar}}, \ and\ \bibinfo {author} {\bibfnamefont {W.}~\bibnamefont
  {Olivares}},\ }\href {\doibase 10.1103/PhysRevE.60.R17} {\bibfield  {journal}
  {\bibinfo  {journal} {Phys. Rev. E}\ }\textbf {\bibinfo {volume} {60}},\
  \bibinfo {pages} {R17} (\bibinfo {year} {1999})}\BibitemShut {NoStop}%
\bibitem [{\citenamefont {Lozada-Cassou}\ and\ \citenamefont
  {Gonz{\'a}lez-Tovar}(2001{\natexlab{a}})}]{Lozada2001}%
  \BibitemOpen
  \bibfield  {author} {\bibinfo {author} {\bibfnamefont {M.}~\bibnamefont
  {Lozada-Cassou}}\ and\ \bibinfo {author} {\bibfnamefont {E.}~\bibnamefont
  {Gonz{\'a}lez-Tovar}},\ }\href {\doibase 10.1006/jcis.2001.7680} {\bibfield
  {journal} {\bibinfo  {journal} {J. Colloid Interface Sci.}\ }\textbf
  {\bibinfo {volume} {239}},\ \bibinfo {pages} {285} (\bibinfo {year}
  {2001}{\natexlab{a}})}\BibitemShut {NoStop}%
\bibitem [{\citenamefont {Lozada-Cassou}\ and\ \citenamefont
  {Gonz{\'a}lez-Tovar}(2001{\natexlab{b}})}]{Lozada2001Erratum}%
  \BibitemOpen
  \bibfield  {author} {\bibinfo {author} {\bibfnamefont {M.}~\bibnamefont
  {Lozada-Cassou}}\ and\ \bibinfo {author} {\bibfnamefont {E.}~\bibnamefont
  {Gonz{\'a}lez-Tovar}},\ }\href {\doibase 10.1006/jcis.2001.7804} {\bibfield
  {journal} {\bibinfo  {journal} {J. Colloid Interface Sci.}\ }\textbf
  {\bibinfo {volume} {240}},\ \bibinfo {pages} {644} (\bibinfo {year}
  {2001}{\natexlab{b}})}\BibitemShut {NoStop}%
\bibitem [{\citenamefont {Manzanilla-Granados}\ \emph
  {et~al.}(2011)\citenamefont {Manzanilla-Granados}, \citenamefont
  {Jimenez-Angeles},\ and\ \citenamefont
  {Lozada-Cassou}}]{ManzanillaGranados:2011}%
  \BibitemOpen
  \bibfield  {author} {\bibinfo {author} {\bibfnamefont {H.}~\bibnamefont
  {Manzanilla-Granados}}, \bibinfo {author} {\bibfnamefont {F.}~\bibnamefont
  {Jimenez-Angeles}}, \ and\ \bibinfo {author} {\bibfnamefont {M.}~\bibnamefont
  {Lozada-Cassou}},\ }\href {\doibase 10.1016/j.colsurfa.2010.11.005}
  {\bibfield  {journal} {\bibinfo  {journal} {Colloids Surf. A}\ }\textbf
  {\bibinfo {volume} {376}},\ \bibinfo {pages} {59} (\bibinfo {year}
  {2011})}\BibitemShut {NoStop}%
\bibitem [{\citenamefont {Carrique}\ \emph {et~al.}(2001)\citenamefont
  {Carrique}, \citenamefont {Arroyo},\ and\ \citenamefont
  {Delgado}}]{Carrique2001157}%
  \BibitemOpen
  \bibfield  {author} {\bibinfo {author} {\bibfnamefont {F.}~\bibnamefont
  {Carrique}}, \bibinfo {author} {\bibfnamefont {F.~J.}\ \bibnamefont
  {Arroyo}}, \ and\ \bibinfo {author} {\bibfnamefont {A.~V.}\ \bibnamefont
  {Delgado}},\ }\href {\doibase 10.1016/S0927-7757(01)00839-1} {\bibfield
  {journal} {\bibinfo  {journal} {Colloids Surf. A}\ }\textbf {\bibinfo
  {volume} {195}},\ \bibinfo {pages} {157} (\bibinfo {year}
  {2001})}\BibitemShut {NoStop}%
\bibitem [{\citenamefont {Carrique}\ \emph {et~al.}(2005)\citenamefont
  {Carrique}, \citenamefont {Cuquejo}, \citenamefont {Arroyo}, \citenamefont
  {Jim{\'e}nez},\ and\ \citenamefont {Delgado}}]{Carrique200543}%
  \BibitemOpen
  \bibfield  {author} {\bibinfo {author} {\bibfnamefont {F.}~\bibnamefont
  {Carrique}}, \bibinfo {author} {\bibfnamefont {J.}~\bibnamefont {Cuquejo}},
  \bibinfo {author} {\bibfnamefont {F.~J.}\ \bibnamefont {Arroyo}}, \bibinfo
  {author} {\bibfnamefont {M.~L.}\ \bibnamefont {Jim{\'e}nez}}, \ and\ \bibinfo
  {author} {\bibfnamefont {A.~V.}\ \bibnamefont {Delgado}},\ }\href {\doibase
  10.1016/j.cis.2005.04.001} {\bibfield  {journal} {\bibinfo  {journal} {Adv.
  Colloid Interface Sci.}\ }\textbf {\bibinfo {volume} {118}},\ \bibinfo
  {pages} {43} (\bibinfo {year} {2005})}\BibitemShut {NoStop}%
\bibitem [{\citenamefont {Chiang}\ \emph {et~al.}(2006)\citenamefont {Chiang},
  \citenamefont {Lee}, \citenamefont {He},\ and\ \citenamefont
  {Hsu}}]{doi:10.1021/jp054969r}%
  \BibitemOpen
  \bibfield  {author} {\bibinfo {author} {\bibfnamefont {C.~P.}\ \bibnamefont
  {Chiang}}, \bibinfo {author} {\bibfnamefont {E.}~\bibnamefont {Lee}},
  \bibinfo {author} {\bibfnamefont {Y.~Y.}\ \bibnamefont {He}}, \ and\ \bibinfo
  {author} {\bibfnamefont {J.~P.}\ \bibnamefont {Hsu}},\ }\href {\doibase
  10.1021/jp054969r} {\bibfield  {journal} {\bibinfo  {journal} {J. Phys. Chem.
  B}\ }\textbf {\bibinfo {volume} {110}},\ \bibinfo {pages} {1490} (\bibinfo
  {year} {2006})},\ \Eprint
  {http://arxiv.org/abs/http://pubs.acs.org/doi/pdf/10.1021/jp054969r}
  {http://pubs.acs.org/doi/pdf/10.1021/jp054969r} \BibitemShut {NoStop}%
\bibitem [{\citenamefont {Ding}\ and\ \citenamefont {Keh}(2001)}]{Ding2001180}%
  \BibitemOpen
  \bibfield  {author} {\bibinfo {author} {\bibfnamefont {J.~M.}\ \bibnamefont
  {Ding}}\ and\ \bibinfo {author} {\bibfnamefont {H.~J.}\ \bibnamefont {Keh}},\
  }\href {\doibase 10.1006/jcis.2000.7383} {\bibfield  {journal} {\bibinfo
  {journal} {J. Colloid Interface Sci.}\ }\textbf {\bibinfo {volume} {236}},\
  \bibinfo {pages} {180} (\bibinfo {year} {2001})}\BibitemShut {NoStop}%
\bibitem [{\citenamefont {Kozak}\ and\ \citenamefont
  {Davis}(1989{\natexlab{a}})}]{Kozak1989497}%
  \BibitemOpen
  \bibfield  {author} {\bibinfo {author} {\bibfnamefont {M.~W.}\ \bibnamefont
  {Kozak}}\ and\ \bibinfo {author} {\bibfnamefont {E.~J.}\ \bibnamefont
  {Davis}},\ }\href {\doibase 10.1016/0021-9797(89)90054-4} {\bibfield
  {journal} {\bibinfo  {journal} {J. Colloid Interface Sci.}\ }\textbf
  {\bibinfo {volume} {127}},\ \bibinfo {pages} {497} (\bibinfo {year}
  {1989}{\natexlab{a}})}\BibitemShut {NoStop}%
\bibitem [{\citenamefont {Kozak}\ and\ \citenamefont
  {Davis}(1989{\natexlab{b}})}]{Kozak1989166}%
  \BibitemOpen
  \bibfield  {author} {\bibinfo {author} {\bibfnamefont {M.~W.}\ \bibnamefont
  {Kozak}}\ and\ \bibinfo {author} {\bibfnamefont {E.~J.}\ \bibnamefont
  {Davis}},\ }\href {\doibase 10.1016/0021-9797(89)90427-X} {\bibfield
  {journal} {\bibinfo  {journal} {J. Colloid Interface Sci.}\ }\textbf
  {\bibinfo {volume} {129}},\ \bibinfo {pages} {166} (\bibinfo {year}
  {1989}{\natexlab{b}})}\BibitemShut {NoStop}%
\bibitem [{\citenamefont {Levine}\ and\ \citenamefont
  {Neale}(1974)}]{Levine1974520}%
  \BibitemOpen
  \bibfield  {author} {\bibinfo {author} {\bibfnamefont {S.}~\bibnamefont
  {Levine}}\ and\ \bibinfo {author} {\bibfnamefont {G.~H.}\ \bibnamefont
  {Neale}},\ }\href {\doibase 10.1016/0021-9797(74)90284-7} {\bibfield
  {journal} {\bibinfo  {journal} {J. Colloid Interface Sci.}\ }\textbf
  {\bibinfo {volume} {47}},\ \bibinfo {pages} {520} (\bibinfo {year}
  {1974})}\BibitemShut {NoStop}%
\bibitem [{\citenamefont {Ohshima}(1997)}]{Ohshima1997481}%
  \BibitemOpen
  \bibfield  {author} {\bibinfo {author} {\bibfnamefont {H.}~\bibnamefont
  {Ohshima}},\ }\href {\doibase 10.1006/jcis.1997.4790} {\bibfield  {journal}
  {\bibinfo  {journal} {J. Colloid Interface Sci.}\ }\textbf {\bibinfo {volume}
  {188}},\ \bibinfo {pages} {481} (\bibinfo {year} {1997})}\BibitemShut
  {NoStop}%
\bibitem [{\citenamefont {Ohshima}(2000)}]{Ohshima2000140}%
  \BibitemOpen
  \bibfield  {author} {\bibinfo {author} {\bibfnamefont {H.}~\bibnamefont
  {Ohshima}},\ }\href {\doibase 10.1006/jcis.2000.6963} {\bibfield  {journal}
  {\bibinfo  {journal} {J. Colloid Interface Sci.}\ }\textbf {\bibinfo {volume}
  {229}},\ \bibinfo {pages} {140} (\bibinfo {year} {2000})}\BibitemShut
  {NoStop}%
\bibitem [{\citenamefont {Shilov}\ \emph {et~al.}(1981)\citenamefont {Shilov},
  \citenamefont {Zharkikh},\ and\ \citenamefont {Borkovskaya}}]{Shilov1981}%
  \BibitemOpen
  \bibfield  {author} {\bibinfo {author} {\bibfnamefont {V.~N.}\ \bibnamefont
  {Shilov}}, \bibinfo {author} {\bibfnamefont {N.~I.}\ \bibnamefont
  {Zharkikh}}, \ and\ \bibinfo {author} {\bibfnamefont {Y.~B.}\ \bibnamefont
  {Borkovskaya}},\ }\href@noop {} {\bibfield  {journal} {\bibinfo  {journal}
  {Colloid J. - USSR}\ }\textbf {\bibinfo {volume} {43}},\ \bibinfo {pages}
  {434} (\bibinfo {year} {1981})}\BibitemShut {NoStop}%
\bibitem [{\citenamefont {Zholkovskiy}\ \emph
  {et~al.}(2007{\natexlab{a}})\citenamefont {Zholkovskiy}, \citenamefont
  {Masliyah}, \citenamefont {Shilov},\ and\ \citenamefont
  {Bhattacharjee}}]{Zholkovskij2007279}%
  \BibitemOpen
  \bibfield  {author} {\bibinfo {author} {\bibfnamefont {E.~K.}\ \bibnamefont
  {Zholkovskiy}}, \bibinfo {author} {\bibfnamefont {J.~H.}\ \bibnamefont
  {Masliyah}}, \bibinfo {author} {\bibfnamefont {V.~N.}\ \bibnamefont
  {Shilov}}, \ and\ \bibinfo {author} {\bibfnamefont {S.}~\bibnamefont
  {Bhattacharjee}},\ }\href {\doibase 10.1016/j.cis.2007.04.025} {\bibfield
  {journal} {\bibinfo  {journal} {Adv. Colloid Interface Sci.}\ }\textbf
  {\bibinfo {volume} {134}},\ \bibinfo {pages} {279} (\bibinfo {year}
  {2007}{\natexlab{a}})}\BibitemShut {NoStop}%
\bibitem [{\citenamefont {Zholkovskiy}\ \emph
  {et~al.}(2007{\natexlab{b}})\citenamefont {Zholkovskiy}, \citenamefont
  {Shilov}, \citenamefont {Masliyah},\ and\ \citenamefont
  {Bondarenko}}]{CJCE:CJCE5450850517}%
  \BibitemOpen
  \bibfield  {author} {\bibinfo {author} {\bibfnamefont {E.~K.}\ \bibnamefont
  {Zholkovskiy}}, \bibinfo {author} {\bibfnamefont {V.~N.}\ \bibnamefont
  {Shilov}}, \bibinfo {author} {\bibfnamefont {J.~H.}\ \bibnamefont
  {Masliyah}}, \ and\ \bibinfo {author} {\bibfnamefont {M.~P.}\ \bibnamefont
  {Bondarenko}},\ }\href {\doibase 10.1002/cjce.5450850517} {\bibfield
  {journal} {\bibinfo  {journal} {Can. J. Chem. Eng.}\ }\textbf {\bibinfo
  {volume} {85}},\ \bibinfo {pages} {701} (\bibinfo {year}
  {2007}{\natexlab{b}})}\BibitemShut {NoStop}%
\bibitem [{\citenamefont {Vissers}\ \emph {et~al.}(2011)\citenamefont
  {Vissers}, \citenamefont {Imhof}, \citenamefont {Carrique}, \citenamefont
  {Delgado},\ and\ \citenamefont {van Blaaderen}}]{Vissers:2011}%
  \BibitemOpen
  \bibfield  {author} {\bibinfo {author} {\bibfnamefont {T.}~\bibnamefont
  {Vissers}}, \bibinfo {author} {\bibfnamefont {A.}~\bibnamefont {Imhof}},
  \bibinfo {author} {\bibfnamefont {F.}~\bibnamefont {Carrique}}, \bibinfo
  {author} {\bibfnamefont {A.~V.}\ \bibnamefont {Delgado}}, \ and\ \bibinfo
  {author} {\bibfnamefont {A.}~\bibnamefont {van Blaaderen}},\ }\href {\doibase
  10.1016/j.jcis.2011.04.113} {\bibfield  {journal} {\bibinfo  {journal} {J.
  Colloid Interface Sci.}\ }\textbf {\bibinfo {volume} {361}},\ \bibinfo
  {pages} {443} (\bibinfo {year} {2011})}\BibitemShut {NoStop}%
\bibitem [{\citenamefont {Hsu}\ \emph {et~al.}(2000)\citenamefont {Hsu},
  \citenamefont {Lee},\ and\ \citenamefont {Yen}}]{Hsu:2000}%
  \BibitemOpen
  \bibfield  {author} {\bibinfo {author} {\bibfnamefont {J.~P.}\ \bibnamefont
  {Hsu}}, \bibinfo {author} {\bibfnamefont {E.}~\bibnamefont {Lee}}, \ and\
  \bibinfo {author} {\bibfnamefont {F.~Y.}\ \bibnamefont {Yen}},\ }\href
  {\doibase 10.1063/1.481203} {\bibfield  {journal} {\bibinfo  {journal} {J.
  Chem. Phys.}\ }\textbf {\bibinfo {volume} {112}},\ \bibinfo {pages} {6404}
  (\bibinfo {year} {2000})}\BibitemShut {NoStop}%
\bibitem [{\citenamefont {Lopez-Garcia}\ \emph {et~al.}(2006)\citenamefont
  {Lopez-Garcia}, \citenamefont {Grosse},\ and\ \citenamefont
  {Horno}}]{LopezGarcia:2006}%
  \BibitemOpen
  \bibfield  {author} {\bibinfo {author} {\bibfnamefont {J.~J.}\ \bibnamefont
  {Lopez-Garcia}}, \bibinfo {author} {\bibfnamefont {C.}~\bibnamefont
  {Grosse}}, \ and\ \bibinfo {author} {\bibfnamefont {J.}~\bibnamefont
  {Horno}},\ }\href {\doibase 10.1016/j.jcis.2006.05.035} {\bibfield  {journal}
  {\bibinfo  {journal} {J. Colloid Interface Sci.}\ }\textbf {\bibinfo {volume}
  {301}},\ \bibinfo {pages} {651} (\bibinfo {year} {2006})}\BibitemShut
  {NoStop}%
\bibitem [{\citenamefont {Ahuali}\ \emph {et~al.}(2009)\citenamefont {Ahuali},
  \citenamefont {Jim{\'e}nez}, \citenamefont {Carrique},\ and\ \citenamefont
  {Delgado}}]{Ahuali:2009}%
  \BibitemOpen
  \bibfield  {author} {\bibinfo {author} {\bibfnamefont {S.}~\bibnamefont
  {Ahuali}}, \bibinfo {author} {\bibfnamefont {M.~L.}\ \bibnamefont
  {Jim{\'e}nez}}, \bibinfo {author} {\bibfnamefont {F.}~\bibnamefont
  {Carrique}}, \ and\ \bibinfo {author} {\bibfnamefont {A.~V.}\ \bibnamefont
  {Delgado}},\ }\href {\doibase 10.1021/la803171f} {\bibfield  {journal}
  {\bibinfo  {journal} {Langmuir}\ }\textbf {\bibinfo {volume} {25}},\ \bibinfo
  {pages} {1986} (\bibinfo {year} {2009})}\BibitemShut {NoStop}%
\bibitem [{\citenamefont {Ennis}\ and\ \citenamefont
  {White}(1997{\natexlab{a}})}]{EnnisWhite:1997}%
  \BibitemOpen
  \bibfield  {author} {\bibinfo {author} {\bibfnamefont {J.}~\bibnamefont
  {Ennis}}\ and\ \bibinfo {author} {\bibfnamefont {L.~R.}\ \bibnamefont
  {White}},\ }\href {\doibase 10.1006/jcis.1996.4565} {\bibfield  {journal}
  {\bibinfo  {journal} {J. Colloid Interface Sci.}\ }\textbf {\bibinfo {volume}
  {185}},\ \bibinfo {pages} {157} (\bibinfo {year}
  {1997}{\natexlab{a}})}\BibitemShut {NoStop}%
\bibitem [{\citenamefont {Ennis}\ and\ \citenamefont
  {White}(1997{\natexlab{b}})}]{EnnisWhite_Corrigendum:1997}%
  \BibitemOpen
  \bibfield  {author} {\bibinfo {author} {\bibfnamefont {J.}~\bibnamefont
  {Ennis}}\ and\ \bibinfo {author} {\bibfnamefont {L.~R.}\ \bibnamefont
  {White}},\ }\href {\doibase 10.1006/jcis.1997.4907} {\bibfield  {journal}
  {\bibinfo  {journal} {J. Colloid Interface Sci.}\ }\textbf {\bibinfo {volume}
  {189}},\ \bibinfo {pages} {382} (\bibinfo {year}
  {1997}{\natexlab{b}})}\BibitemShut {NoStop}%
\bibitem [{\citenamefont {Shugai}\ \emph {et~al.}(1997)\citenamefont {Shugai},
  \citenamefont {Carnie}, \citenamefont {Chan},\ and\ \citenamefont
  {Anderson}}]{ShugaiCarnie:1997}%
  \BibitemOpen
  \bibfield  {author} {\bibinfo {author} {\bibfnamefont {A.~A.}\ \bibnamefont
  {Shugai}}, \bibinfo {author} {\bibfnamefont {S.~L.}\ \bibnamefont {Carnie}},
  \bibinfo {author} {\bibfnamefont {D.~Y.~C.}\ \bibnamefont {Chan}}, \ and\
  \bibinfo {author} {\bibfnamefont {J.~L.}\ \bibnamefont {Anderson}},\ }\href
  {\doibase 10.1006/jcis.1997.4921} {\bibfield  {journal} {\bibinfo  {journal}
  {J. Colloid Interface Sci.}\ }\textbf {\bibinfo {volume} {191}},\ \bibinfo
  {pages} {357} (\bibinfo {year} {1997})}\BibitemShut {NoStop}%
\bibitem [{\citenamefont {Palberg}\ \emph {et~al.}(2012)\citenamefont
  {Palberg}, \citenamefont {K{\"o}ller}, \citenamefont {Sieber}, \citenamefont
  {Schweinfurth}, \citenamefont {Reiber},\ and\ \citenamefont
  {N{\"a}gele}}]{Palberg:2012}%
  \BibitemOpen
  \bibfield  {author} {\bibinfo {author} {\bibfnamefont {T.}~\bibnamefont
  {Palberg}}, \bibinfo {author} {\bibfnamefont {T.}~\bibnamefont {K{\"o}ller}},
  \bibinfo {author} {\bibfnamefont {B.}~\bibnamefont {Sieber}}, \bibinfo
  {author} {\bibfnamefont {H.}~\bibnamefont {Schweinfurth}}, \bibinfo {author}
  {\bibfnamefont {H.}~\bibnamefont {Reiber}}, \ and\ \bibinfo {author}
  {\bibfnamefont {G.}~\bibnamefont {N{\"a}gele}},\ }\href {\doibase
  10.1088/0953-8984/24/46/464109} {\bibfield  {journal} {\bibinfo  {journal}
  {J. Phys.: Condens. Matter}\ }\textbf {\bibinfo {volume} {24}},\ \bibinfo
  {pages} {464109} (\bibinfo {year} {2012})}\BibitemShut {NoStop}%
\bibitem [{\citenamefont {Onsager}\ and\ \citenamefont
  {Fuoss}(1932)}]{Onsager:1932ux}%
  \BibitemOpen
  \bibfield  {author} {\bibinfo {author} {\bibfnamefont {L.}~\bibnamefont
  {Onsager}}\ and\ \bibinfo {author} {\bibfnamefont {R.~M.}\ \bibnamefont
  {Fuoss}},\ }\href {\doibase 10.1021/j150341a001} {\bibfield  {journal}
  {\bibinfo  {journal} {J. Phys. Chem. - US}\ }\textbf {\bibinfo {volume}
  {36}},\ \bibinfo {pages} {2689} (\bibinfo {year} {1932})}\BibitemShut
  {NoStop}%
\bibitem [{\citenamefont {Onsager}\ and\ \citenamefont
  {Kim}(1957)}]{Onsager:1957vv}%
  \BibitemOpen
  \bibfield  {author} {\bibinfo {author} {\bibfnamefont {L.}~\bibnamefont
  {Onsager}}\ and\ \bibinfo {author} {\bibfnamefont {S.~K.}\ \bibnamefont
  {Kim}},\ }\href {\doibase 10.1021/j150548a016} {\bibfield  {journal}
  {\bibinfo  {journal} {J. Phys. Chem. - US}\ }\textbf {\bibinfo {volume}
  {61}},\ \bibinfo {pages} {215} (\bibinfo {year} {1957})}\BibitemShut
  {NoStop}%
\bibitem [{\citenamefont {Onsager}(1945)}]{Onsager:1945we}%
  \BibitemOpen
  \bibfield  {author} {\bibinfo {author} {\bibfnamefont {L.}~\bibnamefont
  {Onsager}},\ }\href {\doibase 10.1111/j.1749-6632.1945.tb36170.x} {\bibfield
  {journal} {\bibinfo  {journal} {Ann. N.Y. Acad. Sci.}\ }\textbf {\bibinfo
  {volume} {46}},\ \bibinfo {pages} {241} (\bibinfo {year} {1945})}\BibitemShut
  {NoStop}%
\bibitem [{\citenamefont {Falkenhagen}\ and\ \citenamefont
  {Vernon}(1932)}]{Falkenhagen:1932us}%
  \BibitemOpen
  \bibfield  {author} {\bibinfo {author} {\bibfnamefont {H.}~\bibnamefont
  {Falkenhagen}}\ and\ \bibinfo {author} {\bibfnamefont {E.~L.}\ \bibnamefont
  {Vernon}},\ }\href@noop {} {\bibfield  {journal} {\bibinfo  {journal} {Phil.
  Mag. S. 7}\ }\textbf {\bibinfo {volume} {14}},\ \bibinfo {pages} {537}
  (\bibinfo {year} {1932})}\BibitemShut {NoStop}%
\bibitem [{\citenamefont {Ebeling}\ \emph {et~al.}(1978)\citenamefont
  {Ebeling}, \citenamefont {Feistel}, \citenamefont {Kelbg},\ and\
  \citenamefont {Sandig}}]{EBELING:1978uo}%
  \BibitemOpen
  \bibfield  {author} {\bibinfo {author} {\bibfnamefont {W.}~\bibnamefont
  {Ebeling}}, \bibinfo {author} {\bibfnamefont {R.}~\bibnamefont {Feistel}},
  \bibinfo {author} {\bibfnamefont {G.}~\bibnamefont {Kelbg}}, \ and\ \bibinfo
  {author} {\bibfnamefont {R.}~\bibnamefont {Sandig}},\ }\href {\doibase
  10.1515/jnet.1978.3.1.11} {\bibfield  {journal} {\bibinfo  {journal} {J.
  Non-Eq. Thermod.}\ }\textbf {\bibinfo {volume} {3}},\ \bibinfo {pages} {11}
  (\bibinfo {year} {1978})}\BibitemShut {NoStop}%
\bibitem [{\citenamefont {Kremp}\ \emph {et~al.}(1983)\citenamefont {Kremp},
  \citenamefont {Ebeling}, \citenamefont {Krienke},\ and\ \citenamefont
  {S{\"a}ndig}}]{KREMP:1983vu}%
  \BibitemOpen
  \bibfield  {author} {\bibinfo {author} {\bibfnamefont {D.}~\bibnamefont
  {Kremp}}, \bibinfo {author} {\bibfnamefont {W.}~\bibnamefont {Ebeling}},
  \bibinfo {author} {\bibfnamefont {H.}~\bibnamefont {Krienke}}, \ and\
  \bibinfo {author} {\bibfnamefont {R.}~\bibnamefont {S{\"a}ndig}},\ }\href
  {\doibase 10.1007/BF01009751} {\bibfield  {journal} {\bibinfo  {journal} {J.
  Stat. Phys.}\ }\textbf {\bibinfo {volume} {33}},\ \bibinfo {pages} {99}
  (\bibinfo {year} {1983})}\BibitemShut {NoStop}%
\bibitem [{\citenamefont {Bernard}\ \emph
  {et~al.}(1992{\natexlab{a}})\citenamefont {Bernard}, \citenamefont {Kunz},
  \citenamefont {Turq},\ and\ \citenamefont {Blum}}]{Bernard:1992tj}%
  \BibitemOpen
  \bibfield  {author} {\bibinfo {author} {\bibfnamefont {O.}~\bibnamefont
  {Bernard}}, \bibinfo {author} {\bibfnamefont {W.}~\bibnamefont {Kunz}},
  \bibinfo {author} {\bibfnamefont {P.}~\bibnamefont {Turq}}, \ and\ \bibinfo
  {author} {\bibfnamefont {L.}~\bibnamefont {Blum}},\ }\href {\doibase
  10.1021/j100188a049} {\bibfield  {journal} {\bibinfo  {journal} {J. Phys.
  Chem. - US}\ }\textbf {\bibinfo {volume} {96}},\ \bibinfo {pages} {3833}
  (\bibinfo {year} {1992}{\natexlab{a}})}\BibitemShut {NoStop}%
\bibitem [{\citenamefont {Durand-Vidal}\ \emph
  {et~al.}(1996{\natexlab{a}})\citenamefont {Durand-Vidal}, \citenamefont
  {Turq}, \citenamefont {Bernard}, \citenamefont {Treiner},\ and\ \citenamefont
  {Blum}}]{DurandVidal:1996tg}%
  \BibitemOpen
  \bibfield  {author} {\bibinfo {author} {\bibfnamefont {S.}~\bibnamefont
  {Durand-Vidal}}, \bibinfo {author} {\bibfnamefont {P.}~\bibnamefont {Turq}},
  \bibinfo {author} {\bibfnamefont {O.}~\bibnamefont {Bernard}}, \bibinfo
  {author} {\bibfnamefont {C.}~\bibnamefont {Treiner}}, \ and\ \bibinfo
  {author} {\bibfnamefont {L.}~\bibnamefont {Blum}},\ }\href {\doibase
  10.1016/0378-4371(96)00083-0} {\bibfield  {journal} {\bibinfo  {journal}
  {Physica A}\ }\textbf {\bibinfo {volume} {231}},\ \bibinfo {pages} {123}
  (\bibinfo {year} {1996}{\natexlab{a}})}\BibitemShut {NoStop}%
\bibitem [{\citenamefont {Durand-Vidal}\ \emph
  {et~al.}(1996{\natexlab{b}})\citenamefont {Durand-Vidal}, \citenamefont
  {Turq},\ and\ \citenamefont {Bernard}}]{DurandVidal:1996vn}%
  \BibitemOpen
  \bibfield  {author} {\bibinfo {author} {\bibfnamefont {S.}~\bibnamefont
  {Durand-Vidal}}, \bibinfo {author} {\bibfnamefont {P.}~\bibnamefont {Turq}},
  \ and\ \bibinfo {author} {\bibfnamefont {O.}~\bibnamefont {Bernard}},\ }\href
  {\doibase 10.1021/jp9613605} {\bibfield  {journal} {\bibinfo  {journal} {J.
  Phys. Chem. - US}\ }\textbf {\bibinfo {volume} {100}},\ \bibinfo {pages}
  {17345} (\bibinfo {year} {1996}{\natexlab{b}})}\BibitemShut {NoStop}%
\bibitem [{\citenamefont {Dufreche}\ \emph
  {et~al.}(2005{\natexlab{a}})\citenamefont {Dufreche}, \citenamefont
  {Bernard}, \citenamefont {Durand-Vidal},\ and\ \citenamefont
  {Turq}}]{Dufreche:2005fj}%
  \BibitemOpen
  \bibfield  {author} {\bibinfo {author} {\bibfnamefont {J.~F.}\ \bibnamefont
  {Dufreche}}, \bibinfo {author} {\bibfnamefont {O.}~\bibnamefont {Bernard}},
  \bibinfo {author} {\bibfnamefont {S.}~\bibnamefont {Durand-Vidal}}, \ and\
  \bibinfo {author} {\bibfnamefont {P.}~\bibnamefont {Turq}},\ }\href {\doibase
  10.1021/jp050387y} {\bibfield  {journal} {\bibinfo  {journal} {J. Phys. Chem.
  B}\ }\textbf {\bibinfo {volume} {109}},\ \bibinfo {pages} {9873} (\bibinfo
  {year} {2005}{\natexlab{a}})}\BibitemShut {NoStop}%
\bibitem [{\citenamefont {Roger}\ \emph {et~al.}(2009)\citenamefont {Roger},
  \citenamefont {Durand-Vidal}, \citenamefont {Bernard},\ and\ \citenamefont
  {Turq}}]{Roger:2009gv}%
  \BibitemOpen
  \bibfield  {author} {\bibinfo {author} {\bibfnamefont {G.~M.}\ \bibnamefont
  {Roger}}, \bibinfo {author} {\bibfnamefont {S.}~\bibnamefont {Durand-Vidal}},
  \bibinfo {author} {\bibfnamefont {O.}~\bibnamefont {Bernard}}, \ and\
  \bibinfo {author} {\bibfnamefont {P.}~\bibnamefont {Turq}},\ }\href {\doibase
  10.1021/jp901916r} {\bibfield  {journal} {\bibinfo  {journal} {J. Phys. Chem.
  B}\ }\textbf {\bibinfo {volume} {113}},\ \bibinfo {pages} {8670} (\bibinfo
  {year} {2009})}\BibitemShut {NoStop}%
\bibitem [{\citenamefont {Blum}\ and\ \citenamefont
  {H{\o}ye}(1977)}]{Blum:1977ep}%
  \BibitemOpen
  \bibfield  {author} {\bibinfo {author} {\bibfnamefont {L.}~\bibnamefont
  {Blum}}\ and\ \bibinfo {author} {\bibfnamefont {J.~S.}\ \bibnamefont
  {H{\o}ye}},\ }\href {\doibase 10.1021/j100528a019} {\bibfield  {journal}
  {\bibinfo  {journal} {J. Phys. Chem.}\ }\textbf {\bibinfo {volume} {81}},\
  \bibinfo {pages} {1311} (\bibinfo {year} {1977})}\BibitemShut {NoStop}%
\bibitem [{\citenamefont {Hiroike}(1977)}]{HIROIKE:1977wm}%
  \BibitemOpen
  \bibfield  {author} {\bibinfo {author} {\bibfnamefont {K.}~\bibnamefont
  {Hiroike}},\ }\href {\doibase 10.1080/00268977700101011} {\bibfield
  {journal} {\bibinfo  {journal} {Mol. Phys.}\ }\textbf {\bibinfo {volume}
  {33}},\ \bibinfo {pages} {1195} (\bibinfo {year} {1977})}\BibitemShut
  {NoStop}%
\bibitem [{\citenamefont {Bernard}\ \emph
  {et~al.}(1992{\natexlab{b}})\citenamefont {Bernard}, \citenamefont {Kunz},
  \citenamefont {Turq},\ and\ \citenamefont {Blum}}]{Bernard:1992we}%
  \BibitemOpen
  \bibfield  {author} {\bibinfo {author} {\bibfnamefont {O.}~\bibnamefont
  {Bernard}}, \bibinfo {author} {\bibfnamefont {W.}~\bibnamefont {Kunz}},
  \bibinfo {author} {\bibfnamefont {P.}~\bibnamefont {Turq}}, \ and\ \bibinfo
  {author} {\bibfnamefont {L.}~\bibnamefont {Blum}},\ }\href {\doibase
  10.1021/j100180a074} {\bibfield  {journal} {\bibinfo  {journal} {J. Phys.
  Chem. - US}\ }\textbf {\bibinfo {volume} {96}},\ \bibinfo {pages} {398}
  (\bibinfo {year} {1992}{\natexlab{b}})}\BibitemShut {NoStop}%
\bibitem [{\citenamefont {Dufreche}\ \emph {et~al.}(2002)\citenamefont
  {Dufreche}, \citenamefont {Bernard},\ and\ \citenamefont
  {Turq}}]{Dufreche:2002wk}%
  \BibitemOpen
  \bibfield  {author} {\bibinfo {author} {\bibfnamefont {J.~F.}\ \bibnamefont
  {Dufreche}}, \bibinfo {author} {\bibfnamefont {O.}~\bibnamefont {Bernard}}, \
  and\ \bibinfo {author} {\bibfnamefont {P.}~\bibnamefont {Turq}},\ }\href
  {\doibase 10.1063/1.1427724} {\bibfield  {journal} {\bibinfo  {journal} {J.
  Chem. Phys.}\ }\textbf {\bibinfo {volume} {116}},\ \bibinfo {pages} {2085}
  (\bibinfo {year} {2002})}\BibitemShut {NoStop}%
\bibitem [{\citenamefont {Dufreche}\ \emph
  {et~al.}(2005{\natexlab{b}})\citenamefont {Dufreche}, \citenamefont
  {Bernard},\ and\ \citenamefont {Turq}}]{DufrecheBrownian:2005}%
  \BibitemOpen
  \bibfield  {author} {\bibinfo {author} {\bibfnamefont {J.~F.}\ \bibnamefont
  {Dufreche}}, \bibinfo {author} {\bibfnamefont {O.}~\bibnamefont {Bernard}}, \
  and\ \bibinfo {author} {\bibfnamefont {P.}~\bibnamefont {Turq}},\ }\href
  {\doibase 10.1016/j.molliq.2004.07.036} {\bibfield  {journal} {\bibinfo
  {journal} {J. Mol. Liq.}\ }\textbf {\bibinfo {volume} {118}},\ \bibinfo
  {pages} {189} (\bibinfo {year} {2005}{\natexlab{b}})}\BibitemShut {NoStop}%
\bibitem [{\citenamefont {Felderhof}(2003)}]{Felderhof:2003ck}%
  \BibitemOpen
  \bibfield  {author} {\bibinfo {author} {\bibfnamefont {B.~U.}\ \bibnamefont
  {Felderhof}},\ }\href {\doibase 10.1063/1.1563604} {\bibfield  {journal}
  {\bibinfo  {journal} {J. Chem. Phys.}\ }\textbf {\bibinfo {volume} {118}},\
  \bibinfo {pages} {8114} (\bibinfo {year} {2003})}\BibitemShut {NoStop}%
\bibitem [{\citenamefont {Dufreche}\ \emph {et~al.}(2003)\citenamefont
  {Dufreche}, \citenamefont {Bernard}, \citenamefont {Jardat},\ and\
  \citenamefont {Turq}}]{Dufreche:2003ku}%
  \BibitemOpen
  \bibfield  {author} {\bibinfo {author} {\bibfnamefont {J.~F.}\ \bibnamefont
  {Dufreche}}, \bibinfo {author} {\bibfnamefont {O.}~\bibnamefont {Bernard}},
  \bibinfo {author} {\bibfnamefont {M.}~\bibnamefont {Jardat}}, \ and\ \bibinfo
  {author} {\bibfnamefont {P.}~\bibnamefont {Turq}},\ }\href {\doibase
  10.1063/1.1563605} {\bibfield  {journal} {\bibinfo  {journal} {J. Chem.
  Phys.}\ }\textbf {\bibinfo {volume} {118}},\ \bibinfo {pages} {8116}
  (\bibinfo {year} {2003})}\BibitemShut {NoStop}%
\bibitem [{\citenamefont {Dufreche}\ \emph {et~al.}(2008)\citenamefont
  {Dufreche}, \citenamefont {Jardat}, \citenamefont {Turq},\ and\ \citenamefont
  {Bagchi}}]{Dufreche:2008fla}%
  \BibitemOpen
  \bibfield  {author} {\bibinfo {author} {\bibfnamefont {J.~F.}\ \bibnamefont
  {Dufreche}}, \bibinfo {author} {\bibfnamefont {M.}~\bibnamefont {Jardat}},
  \bibinfo {author} {\bibfnamefont {P.}~\bibnamefont {Turq}}, \ and\ \bibinfo
  {author} {\bibfnamefont {B.}~\bibnamefont {Bagchi}},\ }\href {\doibase
  10.1021/jp801796g} {\bibfield  {journal} {\bibinfo  {journal} {J. Phys. Chem.
  B}\ }\textbf {\bibinfo {volume} {112}},\ \bibinfo {pages} {10264} (\bibinfo
  {year} {2008})}\BibitemShut {NoStop}%
\bibitem [{\citenamefont {Chandra}\ and\ \citenamefont
  {Bagchi}(1999)}]{Chandra:1999um}%
  \BibitemOpen
  \bibfield  {author} {\bibinfo {author} {\bibfnamefont {A.}~\bibnamefont
  {Chandra}}\ and\ \bibinfo {author} {\bibfnamefont {B.}~\bibnamefont
  {Bagchi}},\ }\href {\doibase 10.1063/1.478876} {\bibfield  {journal}
  {\bibinfo  {journal} {J. Chem. Phys.}\ }\textbf {\bibinfo {volume} {110}},\
  \bibinfo {pages} {10024} (\bibinfo {year} {1999})}\BibitemShut {NoStop}%
\bibitem [{\citenamefont {Chandra}\ and\ \citenamefont
  {Bagchi}(2000{\natexlab{a}})}]{Chandra:2000kj}%
  \BibitemOpen
  \bibfield  {author} {\bibinfo {author} {\bibfnamefont {A.}~\bibnamefont
  {Chandra}}\ and\ \bibinfo {author} {\bibfnamefont {B.}~\bibnamefont
  {Bagchi}},\ }\href {\doibase 10.1021/jp001052d} {\bibfield  {journal}
  {\bibinfo  {journal} {J. Phys. Chem. B}\ }\textbf {\bibinfo {volume} {104}},\
  \bibinfo {pages} {9067} (\bibinfo {year} {2000}{\natexlab{a}})}\BibitemShut
  {NoStop}%
\bibitem [{\citenamefont {Chandra}\ and\ \citenamefont
  {Bagchi}(2000{\natexlab{b}})}]{Chandra:2000um}%
  \BibitemOpen
  \bibfield  {author} {\bibinfo {author} {\bibfnamefont {A.}~\bibnamefont
  {Chandra}}\ and\ \bibinfo {author} {\bibfnamefont {B.}~\bibnamefont
  {Bagchi}},\ }\href {\doibase 10.1063/1.1286963} {\bibfield  {journal}
  {\bibinfo  {journal} {J. Chem. Phys.}\ }\textbf {\bibinfo {volume} {113}},\
  \bibinfo {pages} {3226} (\bibinfo {year} {2000}{\natexlab{b}})}\BibitemShut
  {NoStop}%
\bibitem [{\citenamefont {Attard}(1993)}]{ATTARD:1993vr}%
  \BibitemOpen
  \bibfield  {author} {\bibinfo {author} {\bibfnamefont {P.}~\bibnamefont
  {Attard}},\ }\href {\doibase 10.1103/PhysRevE.48.3604} {\bibfield  {journal}
  {\bibinfo  {journal} {Phys. Rev. E}\ }\textbf {\bibinfo {volume} {48}},\
  \bibinfo {pages} {3604} (\bibinfo {year} {1993})}\BibitemShut {NoStop}%
\bibitem [{\citenamefont {Contreras-Aburto}\ and\ \citenamefont
  {N{\"a}gele}(2013{\natexlab{a}})}]{Contreras_I_Paper:2013}%
  \BibitemOpen
  \bibfield  {author} {\bibinfo {author} {\bibfnamefont {C.}~\bibnamefont
  {Contreras-Aburto}}\ and\ \bibinfo {author} {\bibfnamefont {G.}~\bibnamefont
  {N{\"a}gele}},\ }\href@noop {} {\bibfield  {journal} {\bibinfo  {journal} {J.
  Chem. Phys., accepted}\ } (\bibinfo {year} {2013}{\natexlab{a}})}\BibitemShut
  {NoStop}%
\bibitem [{\citenamefont {Contreras-Aburto}\ and\ \citenamefont
  {N{\"a}gele}(2013{\natexlab{b}})}]{Contreras_II_Paper:2013}%
  \BibitemOpen
  \bibfield  {author} {\bibinfo {author} {\bibfnamefont {C.}~\bibnamefont
  {Contreras-Aburto}}\ and\ \bibinfo {author} {\bibfnamefont {G.}~\bibnamefont
  {N{\"a}gele}},\ }\href@noop {} {\bibfield  {journal} {\bibinfo  {journal} {J.
  Chem. Phys., accepted}\ } (\bibinfo {year} {2013}{\natexlab{b}})}\BibitemShut
  {NoStop}%
\bibitem [{\citenamefont {Contreras-Aburto}\ and\ \citenamefont
  {N{\"a}gele}(2012)}]{Contreras2012visc}%
  \BibitemOpen
  \bibfield  {author} {\bibinfo {author} {\bibfnamefont {C.}~\bibnamefont
  {Contreras-Aburto}}\ and\ \bibinfo {author} {\bibfnamefont {G.}~\bibnamefont
  {N{\"a}gele}},\ }\href {\doibase 10.1088/0953-8984/24/46/464108} {\bibfield
  {journal} {\bibinfo  {journal} {J. Phys.: Condens. Matter}\ }\textbf
  {\bibinfo {volume} {24}},\ \bibinfo {pages} {464108} (\bibinfo {year}
  {2012})}\BibitemShut {NoStop}%
\bibitem [{\citenamefont {N{\"a}gele}\ and\ \citenamefont
  {Dhont}(1998)}]{Nagele:1998tp}%
  \BibitemOpen
  \bibfield  {author} {\bibinfo {author} {\bibfnamefont {G.}~\bibnamefont
  {N{\"a}gele}}\ and\ \bibinfo {author} {\bibfnamefont {J.~K.~G.}\ \bibnamefont
  {Dhont}},\ }\href {\doibase 10.1063/1.476405} {\bibfield  {journal} {\bibinfo
   {journal} {J. Chem. Phys.}\ }\textbf {\bibinfo {volume} {108}},\ \bibinfo
  {pages} {9566} (\bibinfo {year} {1998})}\BibitemShut {NoStop}%
\bibitem [{\citenamefont {N{\"a}gele}\ and\ \citenamefont
  {Bergenholtz}(1998)}]{Nagele:1998tt}%
  \BibitemOpen
  \bibfield  {author} {\bibinfo {author} {\bibfnamefont {G.}~\bibnamefont
  {N{\"a}gele}}\ and\ \bibinfo {author} {\bibfnamefont {J.}~\bibnamefont
  {Bergenholtz}},\ }\href {\doibase 10.1063/1.476428} {\bibfield  {journal}
  {\bibinfo  {journal} {J. Chem. Phys.}\ }\textbf {\bibinfo {volume} {108}},\
  \bibinfo {pages} {9893} (\bibinfo {year} {1998})}\BibitemShut {NoStop}%
\bibitem [{\citenamefont {N{\"a}gele}\ \emph {et~al.}(1999)\citenamefont
  {N{\"a}gele}, \citenamefont {Bergenholtz},\ and\ \citenamefont
  {Dhont}}]{Nagele:1999vh}%
  \BibitemOpen
  \bibfield  {author} {\bibinfo {author} {\bibfnamefont {G.}~\bibnamefont
  {N{\"a}gele}}, \bibinfo {author} {\bibfnamefont {J.}~\bibnamefont
  {Bergenholtz}}, \ and\ \bibinfo {author} {\bibfnamefont {J.~K.~G.}\
  \bibnamefont {Dhont}},\ }\href {\doibase 10.1063/1.478609} {\bibfield
  {journal} {\bibinfo  {journal} {J. Chem. Phys.}\ }\textbf {\bibinfo {volume}
  {110}},\ \bibinfo {pages} {7037} (\bibinfo {year} {1999})}\BibitemShut
  {NoStop}%
\bibitem [{\citenamefont {Miller}(1966)}]{Miller1966}%
  \BibitemOpen
  \bibfield  {author} {\bibinfo {author} {\bibfnamefont {D.~G.}\ \bibnamefont
  {Miller}},\ }\href {\doibase 10.1021/j100880a033} {\bibfield  {journal}
  {\bibinfo  {journal} {J. Phys. Chem. - US}\ }\textbf {\bibinfo {volume}
  {70}},\ \bibinfo {pages} {2639} (\bibinfo {year} {1966})}\BibitemShut
  {NoStop}%
\bibitem [{\citenamefont {Out}\ and\ \citenamefont {Los}(1980)}]{OUT:1980up}%
  \BibitemOpen
  \bibfield  {author} {\bibinfo {author} {\bibfnamefont {D.~J.~P.}\
  \bibnamefont {Out}}\ and\ \bibinfo {author} {\bibfnamefont {J.~M.}\
  \bibnamefont {Los}},\ }\href {\doibase 10.1007/BF00650134} {\bibfield
  {journal} {\bibinfo  {journal} {J. Solution Chem.}\ }\textbf {\bibinfo
  {volume} {9}},\ \bibinfo {pages} {19} (\bibinfo {year} {1980})}\BibitemShut
  {NoStop}%
\bibitem [{\citenamefont {Heinen}\ \emph
  {et~al.}(2011{\natexlab{a}})\citenamefont {Heinen}, \citenamefont {Banchio},\
  and\ \citenamefont {N{\"a}gele}}]{Heinen:2011if}%
  \BibitemOpen
  \bibfield  {author} {\bibinfo {author} {\bibfnamefont {M.}~\bibnamefont
  {Heinen}}, \bibinfo {author} {\bibfnamefont {A.~J.}\ \bibnamefont {Banchio}},
  \ and\ \bibinfo {author} {\bibfnamefont {G.}~\bibnamefont {N{\"a}gele}},\
  }\href {\doibase 10.1063/1.3646962} {\bibfield  {journal} {\bibinfo
  {journal} {J. Chem. Phys.}\ }\textbf {\bibinfo {volume} {135}},\ \bibinfo
  {pages} {154504} (\bibinfo {year} {2011}{\natexlab{a}})}\BibitemShut
  {NoStop}%
\bibitem [{\citenamefont {Bowen}\ and\ \citenamefont
  {Mongruel}(1998)}]{BowenMongruel:1998}%
  \BibitemOpen
  \bibfield  {author} {\bibinfo {author} {\bibfnamefont {W.}~\bibnamefont
  {Bowen}}\ and\ \bibinfo {author} {\bibfnamefont {A.}~\bibnamefont
  {Mongruel}},\ }\href {\doibase 10.1016/S0927-7757(96)03954-4} {\bibfield
  {journal} {\bibinfo  {journal} {Colloids Surf. A}\ }\textbf {\bibinfo
  {volume} {138}},\ \bibinfo {pages} {161} (\bibinfo {year}
  {1998})}\BibitemShut {NoStop}%
\bibitem [{\citenamefont {Yu}\ \emph {et~al.}(2005)\citenamefont {Yu},
  \citenamefont {Tian},\ and\ \citenamefont {Gao}}]{Yu:2005}%
  \BibitemOpen
  \bibfield  {author} {\bibinfo {author} {\bibfnamefont {Y.~X.}\ \bibnamefont
  {Yu}}, \bibinfo {author} {\bibfnamefont {A.~W.}\ \bibnamefont {Tian}}, \ and\
  \bibinfo {author} {\bibfnamefont {G.~H.}\ \bibnamefont {Gao}},\ }\href
  {\doibase 10.1039/B500371G} {\bibfield  {journal} {\bibinfo  {journal} {Phys.
  Chem. Chem. Phys.}\ }\textbf {\bibinfo {volume} {7}},\ \bibinfo {pages}
  {2423} (\bibinfo {year} {2005})}\BibitemShut {NoStop}%
\bibitem [{\citenamefont {Prinsen}\ and\ \citenamefont
  {Odijk}(2007)}]{PrisenOdijk:2007}%
  \BibitemOpen
  \bibfield  {author} {\bibinfo {author} {\bibfnamefont {P.}~\bibnamefont
  {Prinsen}}\ and\ \bibinfo {author} {\bibfnamefont {T.}~\bibnamefont
  {Odijk}},\ }\href {\doibase 10.1063/1.2771160} {\bibfield  {journal}
  {\bibinfo  {journal} {J. Chem. Phys.}\ }\textbf {\bibinfo {volume} {127}},\
  \bibinfo {pages} {115102} (\bibinfo {year} {2007})}\BibitemShut {NoStop}%
\bibitem [{\citenamefont {Heinen}\ \emph
  {et~al.}(2011{\natexlab{b}})\citenamefont {Heinen}, \citenamefont
  {Holmqvist}, \citenamefont {Banchio},\ and\ \citenamefont
  {N{\"a}gele}}]{Heinen_MPBRMSA:2011}%
  \BibitemOpen
  \bibfield  {author} {\bibinfo {author} {\bibfnamefont {M.}~\bibnamefont
  {Heinen}}, \bibinfo {author} {\bibfnamefont {P.}~\bibnamefont {Holmqvist}},
  \bibinfo {author} {\bibfnamefont {A.~J.}\ \bibnamefont {Banchio}}, \ and\
  \bibinfo {author} {\bibfnamefont {G.}~\bibnamefont {N{\"a}gele}},\ }\href
  {\doibase 10.1063/1.3524309} {\bibfield  {journal} {\bibinfo  {journal} {J.
  Chem. Phys.}\ }\textbf {\bibinfo {volume} {134}},\ \bibinfo {pages} {044532}
  (\bibinfo {year} {2011}{\natexlab{b}})}\BibitemShut {NoStop}%
\bibitem [{\citenamefont {Heinen}\ \emph
  {et~al.}(2011{\natexlab{c}})\citenamefont {Heinen}, \citenamefont
  {Holmqvist}, \citenamefont {Banchio},\ and\ \citenamefont
  {N{\"a}gele}}]{Heinen_MPBRMSA_Corrigendum:2011}%
  \BibitemOpen
  \bibfield  {author} {\bibinfo {author} {\bibfnamefont {M.}~\bibnamefont
  {Heinen}}, \bibinfo {author} {\bibfnamefont {P.}~\bibnamefont {Holmqvist}},
  \bibinfo {author} {\bibfnamefont {A.~J.}\ \bibnamefont {Banchio}}, \ and\
  \bibinfo {author} {\bibfnamefont {G.}~\bibnamefont {N{\"a}gele}},\ }\href
  {\doibase 10.1063/1.3570956} {\bibfield  {journal} {\bibinfo  {journal} {J.
  Chem. Phys.}\ }\textbf {\bibinfo {volume} {134}},\ \bibinfo {pages} {129901}
  (\bibinfo {year} {2011}{\natexlab{c}})}\BibitemShut {NoStop}%
\bibitem [{\citenamefont {Kholodenko}\ and\ \citenamefont
  {Douglas}(1995)}]{KholodenkoDouglas:1995}%
  \BibitemOpen
  \bibfield  {author} {\bibinfo {author} {\bibfnamefont {A.~L.}\ \bibnamefont
  {Kholodenko}}\ and\ \bibinfo {author} {\bibfnamefont {J.~F.}\ \bibnamefont
  {Douglas}},\ }\href {\doibase 10.1103/PhysRevE.51.1081} {\bibfield  {journal}
  {\bibinfo  {journal} {Phys. Rev. E}\ }\textbf {\bibinfo {volume} {51}},\
  \bibinfo {pages} {1081} (\bibinfo {year} {1995})}\BibitemShut {NoStop}%
\bibitem [{\citenamefont {Nettesheim}\ \emph {et~al.}(2008)\citenamefont
  {Nettesheim}, \citenamefont {Liberatore}, \citenamefont {Hodgdon},
  \citenamefont {Wagner}, \citenamefont {Kaler},\ and\ \citenamefont
  {Vethamuthu}}]{Nettesheim2008}%
  \BibitemOpen
  \bibfield  {author} {\bibinfo {author} {\bibfnamefont {F.}~\bibnamefont
  {Nettesheim}}, \bibinfo {author} {\bibfnamefont {M.~W.}\ \bibnamefont
  {Liberatore}}, \bibinfo {author} {\bibfnamefont {T.~K.}\ \bibnamefont
  {Hodgdon}}, \bibinfo {author} {\bibfnamefont {N.~J.}\ \bibnamefont {Wagner}},
  \bibinfo {author} {\bibfnamefont {E.~W.}\ \bibnamefont {Kaler}}, \ and\
  \bibinfo {author} {\bibfnamefont {M.}~\bibnamefont {Vethamuthu}},\ }\href
  {\doibase 10.1021/la800271m} {\bibfield  {journal} {\bibinfo  {journal}
  {Langmuir}\ }\textbf {\bibinfo {volume} {24}},\ \bibinfo {pages} {7718}
  (\bibinfo {year} {2008})}\BibitemShut {NoStop}%
\bibitem [{\citenamefont {Gaigalas}\ \emph {et~al.}(1995)\citenamefont
  {Gaigalas}, \citenamefont {Reipa}, \citenamefont {Hubbard}, \citenamefont
  {Edwards},\ and\ \citenamefont {Douglas}}]{Gaigalas1995}%
  \BibitemOpen
  \bibfield  {author} {\bibinfo {author} {\bibfnamefont {A.~K.}\ \bibnamefont
  {Gaigalas}}, \bibinfo {author} {\bibfnamefont {V.}~\bibnamefont {Reipa}},
  \bibinfo {author} {\bibfnamefont {J.~B.}\ \bibnamefont {Hubbard}}, \bibinfo
  {author} {\bibfnamefont {J.}~\bibnamefont {Edwards}}, \ and\ \bibinfo
  {author} {\bibfnamefont {J.}~\bibnamefont {Douglas}},\ }\href {\doibase
  10.1016/0009-2509(94)00465-4} {\bibfield  {journal} {\bibinfo  {journal}
  {Chem. Eng. Sci.}\ }\textbf {\bibinfo {volume} {50}},\ \bibinfo {pages}
  {1107} (\bibinfo {year} {1995})}\BibitemShut {NoStop}%
\bibitem [{\citenamefont {Cohen}\ \emph {et~al.}(1998)\citenamefont {Cohen},
  \citenamefont {Thurston}, \citenamefont {Chamberlin}, \citenamefont
  {Benedek},\ and\ \citenamefont {Carey}}]{Cohen1998}%
  \BibitemOpen
  \bibfield  {author} {\bibinfo {author} {\bibfnamefont {D.~E.}\ \bibnamefont
  {Cohen}}, \bibinfo {author} {\bibfnamefont {G.~M.}\ \bibnamefont {Thurston}},
  \bibinfo {author} {\bibfnamefont {R.~A.}\ \bibnamefont {Chamberlin}},
  \bibinfo {author} {\bibfnamefont {G.~B.}\ \bibnamefont {Benedek}}, \ and\
  \bibinfo {author} {\bibfnamefont {M.~C.}\ \bibnamefont {Carey}},\ }\href
  {\doibase 10.1021/bi980182y} {\bibfield  {journal} {\bibinfo  {journal}
  {Biochemistry}\ }\textbf {\bibinfo {volume} {37}},\ \bibinfo {pages} {14798}
  (\bibinfo {year} {1998})}\BibitemShut {NoStop}%
\bibitem [{\citenamefont {Kleshchanok}\ \emph {et~al.}(2012)\citenamefont
  {Kleshchanok}, \citenamefont {Heinen}, \citenamefont {N{\"a}gele},\ and\
  \citenamefont {Holmqvist}}]{Holmqvist_Gibbsite:2012}%
  \BibitemOpen
  \bibfield  {author} {\bibinfo {author} {\bibfnamefont {D.}~\bibnamefont
  {Kleshchanok}}, \bibinfo {author} {\bibfnamefont {M.}~\bibnamefont {Heinen}},
  \bibinfo {author} {\bibfnamefont {G.}~\bibnamefont {N{\"a}gele}}, \ and\
  \bibinfo {author} {\bibfnamefont {P.}~\bibnamefont {Holmqvist}},\ }\href
  {\doibase 10.1039/c1sm06735d} {\bibfield  {journal} {\bibinfo  {journal}
  {Soft Matter}\ }\textbf {\bibinfo {volume} {8}},\ \bibinfo {pages} {1584}
  (\bibinfo {year} {2012})}\BibitemShut {NoStop}%
\bibitem [{\citenamefont {Kim}\ \emph {et~al.}(2006)\citenamefont {Kim},
  \citenamefont {Nakayama},\ and\ \citenamefont {Yamamoto}}]{Kim:2006}%
  \BibitemOpen
  \bibfield  {author} {\bibinfo {author} {\bibfnamefont {K.}~\bibnamefont
  {Kim}}, \bibinfo {author} {\bibfnamefont {Y.}~\bibnamefont {Nakayama}}, \
  and\ \bibinfo {author} {\bibfnamefont {R.}~\bibnamefont {Yamamoto}},\ }\href
  {\doibase 10.1103/PhysRevLett.96.208302} {\bibfield  {journal} {\bibinfo
  {journal} {Phys. Rev. Lett.}\ }\textbf {\bibinfo {volume} {96}},\ \bibinfo
  {pages} {208302} (\bibinfo {year} {2006})}\BibitemShut {NoStop}%
\bibitem [{\citenamefont {Yamamoto}\ \emph {et~al.}(2009)\citenamefont
  {Yamamoto}, \citenamefont {Nakayama},\ and\ \citenamefont
  {Kim}}]{Yamamoto:2009}%
  \BibitemOpen
  \bibfield  {author} {\bibinfo {author} {\bibfnamefont {E.}~\bibnamefont
  {Yamamoto}}, \bibinfo {author} {\bibfnamefont {Y.}~\bibnamefont {Nakayama}},
  \ and\ \bibinfo {author} {\bibfnamefont {K.}~\bibnamefont {Kim}},\ }\href
  {\doibase 10.1142/S0129183109014515} {\bibfield  {journal} {\bibinfo
  {journal} {Int. J. Mod. Phys. C}\ }\textbf {\bibinfo {volume} {20}},\
  \bibinfo {pages} {1457} (\bibinfo {year} {2009})}\BibitemShut {NoStop}%
\bibitem [{\citenamefont {Araki}\ and\ \citenamefont
  {Tanaka}(2008)}]{Araki_Tanaka:2008}%
  \BibitemOpen
  \bibfield  {author} {\bibinfo {author} {\bibfnamefont {T.}~\bibnamefont
  {Araki}}\ and\ \bibinfo {author} {\bibfnamefont {H.}~\bibnamefont {Tanaka}},\
  }\href {\doibase 10.1209/0295-5075/82/18004} {\bibfield  {journal} {\bibinfo
  {journal} {Europhys. Lett.}\ }\textbf {\bibinfo {volume} {82}},\ \bibinfo
  {pages} {18004} (\bibinfo {year} {2008})}\BibitemShut {NoStop}%
\bibitem [{\citenamefont {Giupponi}\ and\ \citenamefont
  {Pagonabarraga}(2011)}]{Giupponi:2011}%
  \BibitemOpen
  \bibfield  {author} {\bibinfo {author} {\bibfnamefont {G.}~\bibnamefont
  {Giupponi}}\ and\ \bibinfo {author} {\bibfnamefont {I.}~\bibnamefont
  {Pagonabarraga}},\ }\href {\doibase 10.1103/PhysRevLett.106.248304}
  {\bibfield  {journal} {\bibinfo  {journal} {Phys. Rev. Lett.}\ }\textbf
  {\bibinfo {volume} {106}},\ \bibinfo {pages} {248304} (\bibinfo {year}
  {2011})}\BibitemShut {NoStop}%
\bibitem [{\citenamefont {Abade}\ \emph
  {et~al.}(2010{\natexlab{a}})\citenamefont {Abade}, \citenamefont {Cichocki},
  \citenamefont {Ekiel-Jezewska}, \citenamefont {N{\"a}gele},\ and\
  \citenamefont {Wajnryb}}]{Abade:2010gt}%
  \BibitemOpen
  \bibfield  {author} {\bibinfo {author} {\bibfnamefont {G.~C.}\ \bibnamefont
  {Abade}}, \bibinfo {author} {\bibfnamefont {B.}~\bibnamefont {Cichocki}},
  \bibinfo {author} {\bibfnamefont {M.~L.}\ \bibnamefont {Ekiel-Jezewska}},
  \bibinfo {author} {\bibfnamefont {G.}~\bibnamefont {N{\"a}gele}}, \ and\
  \bibinfo {author} {\bibfnamefont {E.}~\bibnamefont {Wajnryb}},\ }\href
  {\doibase 10.1063/1.3274663} {\bibfield  {journal} {\bibinfo  {journal} {J.
  Chem. Phys.}\ }\textbf {\bibinfo {volume} {132}},\ \bibinfo {pages} {014503}
  (\bibinfo {year} {2010}{\natexlab{a}})}\BibitemShut {NoStop}%
\bibitem [{\citenamefont {Reuland}\ \emph {et~al.}(1978)\citenamefont
  {Reuland}, \citenamefont {Felderhof},\ and\ \citenamefont
  {Jones}}]{1978PhyA...93..465R}%
  \BibitemOpen
  \bibfield  {author} {\bibinfo {author} {\bibfnamefont {P.}~\bibnamefont
  {Reuland}}, \bibinfo {author} {\bibfnamefont {B.~U.}\ \bibnamefont
  {Felderhof}}, \ and\ \bibinfo {author} {\bibfnamefont {R.~B.}\ \bibnamefont
  {Jones}},\ }\href {\doibase 10.1016/0378-4371(78)90167-X} {\bibfield
  {journal} {\bibinfo  {journal} {Physica A}\ }\textbf {\bibinfo {volume}
  {93}},\ \bibinfo {pages} {465} (\bibinfo {year} {1978})}\BibitemShut
  {NoStop}%
\bibitem [{\citenamefont {Lobaskin}\ \emph {et~al.}(2004)\citenamefont
  {Lobaskin}, \citenamefont {D{\"u}nweg},\ and\ \citenamefont
  {Holm}}]{Lobaskin:2004}%
  \BibitemOpen
  \bibfield  {author} {\bibinfo {author} {\bibfnamefont {V.}~\bibnamefont
  {Lobaskin}}, \bibinfo {author} {\bibfnamefont {B.}~\bibnamefont
  {D{\"u}nweg}}, \ and\ \bibinfo {author} {\bibfnamefont {C.}~\bibnamefont
  {Holm}},\ }\href {\doibase 10.1088/0953-8984/16/38/021} {\bibfield  {journal}
  {\bibinfo  {journal} {J. Phys.: Condens. Matter}\ }\textbf {\bibinfo {volume}
  {16}},\ \bibinfo {pages} {S4063} (\bibinfo {year} {2004})}\BibitemShut
  {NoStop}%
\bibitem [{\citenamefont {Lobaskin}\ \emph {et~al.}(2007)\citenamefont
  {Lobaskin}, \citenamefont {D{\"u}nweg}, \citenamefont {Medebach},
  \citenamefont {Palberg},\ and\ \citenamefont {Holm}}]{Lobaskin_PRL:2007}%
  \BibitemOpen
  \bibfield  {author} {\bibinfo {author} {\bibfnamefont {V.}~\bibnamefont
  {Lobaskin}}, \bibinfo {author} {\bibfnamefont {B.}~\bibnamefont
  {D{\"u}nweg}}, \bibinfo {author} {\bibfnamefont {B.}~\bibnamefont
  {Medebach}}, \bibinfo {author} {\bibfnamefont {T.}~\bibnamefont {Palberg}}, \
  and\ \bibinfo {author} {\bibfnamefont {C.}~\bibnamefont {Holm}},\ }\href
  {\doibase 10.1103/PhysRevLett.98.176105} {\bibfield  {journal} {\bibinfo
  {journal} {Phys. Rev. Lett.}\ }\textbf {\bibinfo {volume} {98}},\ \bibinfo
  {pages} {176105} (\bibinfo {year} {2007})}\BibitemShut {NoStop}%
\bibitem [{\citenamefont {Chatterji}\ and\ \citenamefont
  {Horbach}(2010)}]{Chatterji:2011}%
  \BibitemOpen
  \bibfield  {author} {\bibinfo {author} {\bibfnamefont {A.}~\bibnamefont
  {Chatterji}}\ and\ \bibinfo {author} {\bibfnamefont {J.}~\bibnamefont
  {Horbach}},\ }\href {\doibase 10.1088/0953-8984/22/49/494102} {\bibfield
  {journal} {\bibinfo  {journal} {J. Phys.: Condens. Matter}\ }\textbf
  {\bibinfo {volume} {22}},\ \bibinfo {pages} {494102} (\bibinfo {year}
  {2010})}\BibitemShut {NoStop}%
\bibitem [{\citenamefont {Schmitz}\ and\ \citenamefont
  {D{\"u}nweg}(2012)}]{Duenweg:2012}%
  \BibitemOpen
  \bibfield  {author} {\bibinfo {author} {\bibfnamefont {R.}~\bibnamefont
  {Schmitz}}\ and\ \bibinfo {author} {\bibfnamefont {B.}~\bibnamefont
  {D{\"u}nweg}},\ }\href {\doibase 10.1088/0953-8984/24/46/464111} {\bibfield
  {journal} {\bibinfo  {journal} {J. Phys.: Condens. Matter}\ }\textbf
  {\bibinfo {volume} {24}},\ \bibinfo {pages} {464111} (\bibinfo {year}
  {2012})}\BibitemShut {NoStop}%
\bibitem [{\citenamefont {D{\"u}nweg}\ \emph {et~al.}(2008)\citenamefont
  {D{\"u}nweg}, \citenamefont {Lobaskin}, \citenamefont
  {Seethalakshmy-Hariharan},\ and\ \citenamefont {Holm}}]{Duenweg:2008}%
  \BibitemOpen
  \bibfield  {author} {\bibinfo {author} {\bibfnamefont {B.}~\bibnamefont
  {D{\"u}nweg}}, \bibinfo {author} {\bibfnamefont {V.}~\bibnamefont
  {Lobaskin}}, \bibinfo {author} {\bibfnamefont {K.}~\bibnamefont
  {Seethalakshmy-Hariharan}}, \ and\ \bibinfo {author} {\bibfnamefont
  {C.}~\bibnamefont {Holm}},\ }\href {\doibase 10.1088/0953-8984/20/40/404214}
  {\bibfield  {journal} {\bibinfo  {journal} {J. Phys.: Condens. Matter}\
  }\textbf {\bibinfo {volume} {20}},\ \bibinfo {pages} {404214} (\bibinfo
  {year} {2008})}\BibitemShut {NoStop}%
\bibitem [{\citenamefont {McPhie}\ and\ \citenamefont
  {N{\"a}gele}(2007)}]{McPhie:2007ee}%
  \BibitemOpen
  \bibfield  {author} {\bibinfo {author} {\bibfnamefont {M.~G.}\ \bibnamefont
  {McPhie}}\ and\ \bibinfo {author} {\bibfnamefont {G.}~\bibnamefont
  {N{\"a}gele}},\ }\href {\doibase 10.1063/1.2753839} {\bibfield  {journal}
  {\bibinfo  {journal} {J. Chem. Phys.}\ }\textbf {\bibinfo {volume} {127}},\
  \bibinfo {pages} {034906} (\bibinfo {year} {2007})}\BibitemShut {NoStop}%
\bibitem [{\citenamefont {McPhie}\ and\ \citenamefont
  {N{\"a}gele}(2004)}]{McPhie:2004gy}%
  \BibitemOpen
  \bibfield  {author} {\bibinfo {author} {\bibfnamefont {M.~G.}\ \bibnamefont
  {McPhie}}\ and\ \bibinfo {author} {\bibfnamefont {G.}~\bibnamefont
  {N{\"a}gele}},\ }\href {\doibase 10.1088/0953-8984/16/38/018} {\bibfield
  {journal} {\bibinfo  {journal} {J Phys-Condens Mat}\ }\textbf {\bibinfo
  {volume} {16}},\ \bibinfo {pages} {S4021} (\bibinfo {year}
  {2004})}\BibitemShut {NoStop}%
\bibitem [{\citenamefont {Banchio}\ and\ \citenamefont
  {N{\"a}gele}(2008)}]{Banchio:2008gt}%
  \BibitemOpen
  \bibfield  {author} {\bibinfo {author} {\bibfnamefont {A.~J.}\ \bibnamefont
  {Banchio}}\ and\ \bibinfo {author} {\bibfnamefont {G.}~\bibnamefont
  {N{\"a}gele}},\ }\href {\doibase 10.1063/1.2868773} {\bibfield  {journal}
  {\bibinfo  {journal} {J. Chem. Phys.}\ }\textbf {\bibinfo {volume} {128}},\
  \bibinfo {pages} {104903} (\bibinfo {year} {2008})}\BibitemShut {NoStop}%
\bibitem [{\citenamefont {Heinen}\ \emph {et~al.}(2010)\citenamefont {Heinen},
  \citenamefont {Holmqvist}, \citenamefont {Banchio},\ and\ \citenamefont
  {N{\"a}gele}}]{Heinen_Appl_cryt:2010}%
  \BibitemOpen
  \bibfield  {author} {\bibinfo {author} {\bibfnamefont {M.}~\bibnamefont
  {Heinen}}, \bibinfo {author} {\bibfnamefont {P.}~\bibnamefont {Holmqvist}},
  \bibinfo {author} {\bibfnamefont {A.~J.}\ \bibnamefont {Banchio}}, \ and\
  \bibinfo {author} {\bibfnamefont {G.}~\bibnamefont {N{\"a}gele}},\ }\href
  {\doibase 10.1107/S002188981002724X} {\bibfield  {journal} {\bibinfo
  {journal} {J. Appl. Cryst.}\ }\textbf {\bibinfo {volume} {43}},\ \bibinfo
  {pages} {970} (\bibinfo {year} {2010})}\BibitemShut {NoStop}%
\bibitem [{\citenamefont {Westermeier}\ \emph {et~al.}(2012)\citenamefont
  {Westermeier}, \citenamefont {Fischer}, \citenamefont {Roseker},
  \citenamefont {Gr{\"u}bel}, \citenamefont {N{\"a}gele},\ and\ \citenamefont
  {Heinen}}]{Westermeier:2012}%
  \BibitemOpen
  \bibfield  {author} {\bibinfo {author} {\bibfnamefont {F.}~\bibnamefont
  {Westermeier}}, \bibinfo {author} {\bibfnamefont {B.}~\bibnamefont
  {Fischer}}, \bibinfo {author} {\bibfnamefont {W.}~\bibnamefont {Roseker}},
  \bibinfo {author} {\bibfnamefont {G.}~\bibnamefont {Gr{\"u}bel}}, \bibinfo
  {author} {\bibfnamefont {G.}~\bibnamefont {N{\"a}gele}}, \ and\ \bibinfo
  {author} {\bibfnamefont {M.}~\bibnamefont {Heinen}},\ }\href {\doibase
  10.1063/1.4751544} {\bibfield  {journal} {\bibinfo  {journal} {J. Chem.
  Phys.}\ }\textbf {\bibinfo {volume} {137}},\ \bibinfo {pages} {114504}
  (\bibinfo {year} {2012})}\BibitemShut {NoStop}%
\bibitem [{\citenamefont {Banchio}\ \emph {et~al.}(2013)\citenamefont
  {Banchio}, \citenamefont {Holmqvist}, \citenamefont {Heinen},\ and\
  \citenamefont {N{\"a}gele}}]{Banchio_Scaling:2013}%
  \BibitemOpen
  \bibfield  {author} {\bibinfo {author} {\bibfnamefont {A.~J.}\ \bibnamefont
  {Banchio}}, \bibinfo {author} {\bibfnamefont {P.}~\bibnamefont {Holmqvist}},
  \bibinfo {author} {\bibfnamefont {M.}~\bibnamefont {Heinen}}, \ and\ \bibinfo
  {author} {\bibfnamefont {G.}~\bibnamefont {N{\"a}gele}},\ }\href@noop {} {}
  (\bibinfo {year} {2013}),\ \bibinfo {note} {submitted}\BibitemShut {NoStop}%
\bibitem [{\citenamefont {Cichocki}\ \emph {et~al.}(1994)\citenamefont
  {Cichocki}, \citenamefont {Felderhof}, \citenamefont {Hinsen}, \citenamefont
  {Wajnryb},\ and\ \citenamefont
  {Blawzdziewicz}}]{Cichocki_HYDROMULTIPOLE:1994}%
  \BibitemOpen
  \bibfield  {author} {\bibinfo {author} {\bibfnamefont {B.}~\bibnamefont
  {Cichocki}}, \bibinfo {author} {\bibfnamefont {B.~U.}\ \bibnamefont
  {Felderhof}}, \bibinfo {author} {\bibfnamefont {K.}~\bibnamefont {Hinsen}},
  \bibinfo {author} {\bibfnamefont {E.}~\bibnamefont {Wajnryb}}, \ and\
  \bibinfo {author} {\bibfnamefont {J.}~\bibnamefont {Blawzdziewicz}},\ }\href
  {\doibase 10.1063/1.466366} {\bibfield  {journal} {\bibinfo  {journal} {J.
  Chem. Phys.}\ }\textbf {\bibinfo {volume} {100}},\ \bibinfo {pages} {3780}
  (\bibinfo {year} {1994})}\BibitemShut {NoStop}%
\bibitem [{\citenamefont {Abade}\ \emph
  {et~al.}(2010{\natexlab{b}})\citenamefont {Abade}, \citenamefont {Cichocki},
  \citenamefont {Ekiel-Jezewska}, \citenamefont {N{\"a}gele},\ and\
  \citenamefont {Wajnryb}}]{Abade_JPCM_2010}%
  \BibitemOpen
  \bibfield  {author} {\bibinfo {author} {\bibfnamefont {G.~C.}\ \bibnamefont
  {Abade}}, \bibinfo {author} {\bibfnamefont {B.}~\bibnamefont {Cichocki}},
  \bibinfo {author} {\bibfnamefont {M.~L.}\ \bibnamefont {Ekiel-Jezewska}},
  \bibinfo {author} {\bibfnamefont {G.}~\bibnamefont {N{\"a}gele}}, \ and\
  \bibinfo {author} {\bibfnamefont {E.}~\bibnamefont {Wajnryb}},\ }\href
  {\doibase 10.1088/0953-8984/22/32/322101} {\bibfield  {journal} {\bibinfo
  {journal} {J. Phys.: Condens. Matter}\ }\textbf {\bibinfo {volume} {22}},\
  \bibinfo {pages} {322101} (\bibinfo {year} {2010}{\natexlab{b}})}\BibitemShut
  {NoStop}%
\bibitem [{\citenamefont {Abade}\ \emph
  {et~al.}(2010{\natexlab{c}})\citenamefont {Abade}, \citenamefont {Cichocki},
  \citenamefont {Ekiel-Jezewska}, \citenamefont {N{\"a}gele},\ and\
  \citenamefont {Wajnryb}}]{AbadeViscosityJCP:2010}%
  \BibitemOpen
  \bibfield  {author} {\bibinfo {author} {\bibfnamefont {G.~C.}\ \bibnamefont
  {Abade}}, \bibinfo {author} {\bibfnamefont {B.}~\bibnamefont {Cichocki}},
  \bibinfo {author} {\bibfnamefont {M.~L.}\ \bibnamefont {Ekiel-Jezewska}},
  \bibinfo {author} {\bibfnamefont {G.}~\bibnamefont {N{\"a}gele}}, \ and\
  \bibinfo {author} {\bibfnamefont {E.}~\bibnamefont {Wajnryb}},\ }\href
  {\doibase 10.1063/1.3474804} {\bibfield  {journal} {\bibinfo  {journal} {J.
  Chem. Phys.}\ }\textbf {\bibinfo {volume} {133}},\ \bibinfo {pages} {084906}
  (\bibinfo {year} {2010}{\natexlab{c}})}\BibitemShut {NoStop}%
\bibitem [{\citenamefont {Abade}\ \emph {et~al.}(2012)\citenamefont {Abade},
  \citenamefont {Cichocki}, \citenamefont {Ekiel-Jezewska}, \citenamefont
  {N{\"a}gele},\ and\ \citenamefont {Wajnryb}}]{AbadeCoreshellJCP:2012}%
  \BibitemOpen
  \bibfield  {author} {\bibinfo {author} {\bibfnamefont {G.~C.}\ \bibnamefont
  {Abade}}, \bibinfo {author} {\bibfnamefont {B.}~\bibnamefont {Cichocki}},
  \bibinfo {author} {\bibfnamefont {M.~L.}\ \bibnamefont {Ekiel-Jezewska}},
  \bibinfo {author} {\bibfnamefont {G.}~\bibnamefont {N{\"a}gele}}, \ and\
  \bibinfo {author} {\bibfnamefont {E.}~\bibnamefont {Wajnryb}},\ }\href
  {\doibase 10.1063/1.3689322} {\bibfield  {journal} {\bibinfo  {journal} {J.
  Chem. Phys.}\ }\textbf {\bibinfo {volume} {136}},\ \bibinfo {pages} {104902}
  (\bibinfo {year} {2012})}\BibitemShut {NoStop}%
\bibitem [{\citenamefont {Ohshima}(2009)}]{Ohshima:2009}%
  \BibitemOpen
  \bibfield  {author} {\bibinfo {author} {\bibfnamefont {H.}~\bibnamefont
  {Ohshima}},\ }\href {\doibase 10.1016/j.colsurfa.2008.11.017} {\bibfield
  {journal} {\bibinfo  {journal} {Colloids Surf. A}\ }\textbf {\bibinfo
  {volume} {347}},\ \bibinfo {pages} {33} (\bibinfo {year} {2009})}\BibitemShut
  {NoStop}%
\bibitem [{\citenamefont {Makuch}\ and\ \citenamefont
  {Cichocki}(2012)}]{Makuch:2012}%
  \BibitemOpen
  \bibfield  {author} {\bibinfo {author} {\bibfnamefont {K.}~\bibnamefont
  {Makuch}}\ and\ \bibinfo {author} {\bibfnamefont {B.}~\bibnamefont
  {Cichocki}},\ }\href {\doibase 10.1063/1.4764303} {\bibfield  {journal}
  {\bibinfo  {journal} {J. Chem. Phys.}\ }\textbf {\bibinfo {volume} {137}},\
  \bibinfo {pages} {184902} (\bibinfo {year} {2012})}\BibitemShut {NoStop}%
\bibitem [{\citenamefont {Russel}(2009)}]{RusselViscosity:2009}%
  \BibitemOpen
  \bibfield  {author} {\bibinfo {author} {\bibfnamefont {W.~B.}\ \bibnamefont
  {Russel}},\ }\href {\doibase 10.1021/ie800385m} {\bibfield  {journal}
  {\bibinfo  {journal} {Ind. Eng. Chem. Res.}\ }\textbf {\bibinfo {volume}
  {48}},\ \bibinfo {pages} {2380} (\bibinfo {year} {2009})}\BibitemShut
  {NoStop}%
\bibitem [{\citenamefont {Holmqvist}\ and\ \citenamefont
  {N{\"a}gele}(2010)}]{Holmqvist_PRL:2010}%
  \BibitemOpen
  \bibfield  {author} {\bibinfo {author} {\bibfnamefont {P.}~\bibnamefont
  {Holmqvist}}\ and\ \bibinfo {author} {\bibfnamefont {G.}~\bibnamefont
  {N{\"a}gele}},\ }\href {\doibase 10.1103/PhysRevLett.104.058301} {\bibfield
  {journal} {\bibinfo  {journal} {Phys. Rev. Lett.}\ }\textbf {\bibinfo
  {volume} {104}},\ \bibinfo {pages} {058301} (\bibinfo {year}
  {2010})}\BibitemShut {NoStop}%
\bibitem [{\citenamefont {Segr\`{e}}\ and\ \citenamefont
  {Pusey}(1996)}]{SegrePusey_PRL:1996}%
  \BibitemOpen
  \bibfield  {author} {\bibinfo {author} {\bibfnamefont {P.~N.}\ \bibnamefont
  {Segr\`{e}}}\ and\ \bibinfo {author} {\bibfnamefont {P.~N.}\ \bibnamefont
  {Pusey}},\ }\href {\doibase 10.1103/PhysRevLett.77.771} {\bibfield  {journal}
  {\bibinfo  {journal} {Phys. Rev. Lett.}\ }\textbf {\bibinfo {volume} {77}},\
  \bibinfo {pages} {771} (\bibinfo {year} {1996})}\BibitemShut {NoStop}%
\bibitem [{\citenamefont {Fuchs}\ and\ \citenamefont {Mayr}(1999)}]{Fuchs1999}%
  \BibitemOpen
  \bibfield  {author} {\bibinfo {author} {\bibfnamefont {M.}~\bibnamefont
  {Fuchs}}\ and\ \bibinfo {author} {\bibfnamefont {M.~R.}\ \bibnamefont
  {Mayr}},\ }\href {\doibase 10.1103/PhysRevE.60.5742} {\bibfield  {journal}
  {\bibinfo  {journal} {Phys. Rev. E}\ }\textbf {\bibinfo {volume} {60}},\
  \bibinfo {pages} {5742} (\bibinfo {year} {1999})}\BibitemShut {NoStop}%
\bibitem [{\citenamefont {Lurio}\ \emph {et~al.}(2000)\citenamefont {Lurio},
  \citenamefont {Lumma}, \citenamefont {Sandy}, \citenamefont {Borthwick},
  \citenamefont {Falus}, \citenamefont {Mochrie}, \citenamefont {Pelletier},
  \citenamefont {Sutton}, \citenamefont {Regan}, \citenamefont {Malik},\ and\
  \citenamefont {Stephenson}}]{Lurio_PRL:2000}%
  \BibitemOpen
  \bibfield  {author} {\bibinfo {author} {\bibfnamefont {L.~B.}\ \bibnamefont
  {Lurio}}, \bibinfo {author} {\bibfnamefont {D.}~\bibnamefont {Lumma}},
  \bibinfo {author} {\bibfnamefont {A.~R.}\ \bibnamefont {Sandy}}, \bibinfo
  {author} {\bibfnamefont {M.~A.}\ \bibnamefont {Borthwick}}, \bibinfo {author}
  {\bibfnamefont {P.}~\bibnamefont {Falus}}, \bibinfo {author} {\bibfnamefont
  {S.~G.~J.}\ \bibnamefont {Mochrie}}, \bibinfo {author} {\bibfnamefont
  {J.~F.}\ \bibnamefont {Pelletier}}, \bibinfo {author} {\bibfnamefont
  {M.}~\bibnamefont {Sutton}}, \bibinfo {author} {\bibfnamefont
  {L.}~\bibnamefont {Regan}}, \bibinfo {author} {\bibfnamefont
  {A.}~\bibnamefont {Malik}}, \ and\ \bibinfo {author} {\bibfnamefont {G.~B.}\
  \bibnamefont {Stephenson}},\ }\href {\doibase 10.1103/PhysRevLett.84.785}
  {\bibfield  {journal} {\bibinfo  {journal} {Phys. Rev. Lett.}\ }\textbf
  {\bibinfo {volume} {84}},\ \bibinfo {pages} {785} (\bibinfo {year}
  {2000})}\BibitemShut {NoStop}%
\bibitem [{\citenamefont {Martinez}\ \emph {et~al.}(2011)\citenamefont
  {Martinez}, \citenamefont {Thijssen}, \citenamefont {Zontone}, \citenamefont
  {van Megen},\ and\ \citenamefont {Bryant}}]{MartinezScaling:2011}%
  \BibitemOpen
  \bibfield  {author} {\bibinfo {author} {\bibfnamefont {V.~A.}\ \bibnamefont
  {Martinez}}, \bibinfo {author} {\bibfnamefont {J.~H.~J.}\ \bibnamefont
  {Thijssen}}, \bibinfo {author} {\bibfnamefont {F.}~\bibnamefont {Zontone}},
  \bibinfo {author} {\bibfnamefont {W.}~\bibnamefont {van Megen}}, \ and\
  \bibinfo {author} {\bibfnamefont {G.}~\bibnamefont {Bryant}},\ }\href
  {\doibase 10.1063/1.3525101} {\bibfield  {journal} {\bibinfo  {journal} {J.
  Chem. Phys.}\ }\textbf {\bibinfo {volume} {134}},\ \bibinfo {pages} {054505}
  (\bibinfo {year} {2011})}\BibitemShut {NoStop}%
\bibitem [{\citenamefont {Holmqvist}\ \emph {et~al.}(2012)\citenamefont
  {Holmqvist}, \citenamefont {Mohanty}, \citenamefont {N{\"a}gele},
  \citenamefont {Schurtenberger},\ and\ \citenamefont
  {Heinen}}]{Holmqvist_PRL:2012}%
  \BibitemOpen
  \bibfield  {author} {\bibinfo {author} {\bibfnamefont {P.}~\bibnamefont
  {Holmqvist}}, \bibinfo {author} {\bibfnamefont {P.~S.}\ \bibnamefont
  {Mohanty}}, \bibinfo {author} {\bibfnamefont {G.}~\bibnamefont {N{\"a}gele}},
  \bibinfo {author} {\bibfnamefont {P.}~\bibnamefont {Schurtenberger}}, \ and\
  \bibinfo {author} {\bibfnamefont {M.}~\bibnamefont {Heinen}},\ }\href
  {\doibase 10.1103/PhysRevLett.109.048302} {\bibfield  {journal} {\bibinfo
  {journal} {Phys. Rev. Lett.}\ }\textbf {\bibinfo {volume} {109}},\ \bibinfo
  {pages} {048302} (\bibinfo {year} {2012})}\BibitemShut {NoStop}%
\bibitem [{\citenamefont {Denton}(2003)}]{Denton:2003}%
  \BibitemOpen
  \bibfield  {author} {\bibinfo {author} {\bibfnamefont {A.~R.}\ \bibnamefont
  {Denton}},\ }\href {\doibase 10.1103/PhysRevE.67.011804} {\bibfield
  {journal} {\bibinfo  {journal} {Phys. Rev. E}\ }\textbf {\bibinfo {volume}
  {67}},\ \bibinfo {pages} {011804} (\bibinfo {year} {2003})}\BibitemShut
  {NoStop}%
\bibitem [{\citenamefont {Likos}(2011)}]{Likos_MicrogelReview:2011}%
  \BibitemOpen
  \bibfield  {author} {\bibinfo {author} {\bibfnamefont {C.~N.}\ \bibnamefont
  {Likos}},\ }\enquote {\bibinfo {title} {Structure and thermodynamics of ionic
  microgels, in: Microgel suspensions: Fundamentals and applications},}\ \
  (\bibinfo  {publisher} {Wiley-VCH Verlag GmbH \& Co. KGaA},\ \bibinfo {year}
  {2011})\ Chap.~\bibinfo {chapter} {7}\BibitemShut {NoStop}%
\end{thebibliography}%

\end{document}